# Sub-Nanometer Channels Embedded in Two-Dimensional Materials

Yimo Han[1]*, Ming-Yang Li[2,3]*, Gang-Seob Jung[4]*, Mark A. Marsalis[5], Zhao Qin[4], Markus J. Buehler[4], Lain-Jong Li[2]†, David A. Muller[1,6]†

[1.] School of Applied & Engineering Physics, Cornell University, Ithaca, NY, 14850, USA

[2.] Physical Science and Engineering Division, King Abdullah University of Science and Technology, Thuwal, 23955-6900, Kingdom of Saudi Arabia

[3.] Research Center for Applied Sciences, Academia Sinica, Taipei, 10617, Taiwan

[4.] Department of Civil and Environmental Engineering, MIT, Cambridge, MA, 02139, USA

[5.] Department of Physics, Texas Tech University, Lubbock, TX, 79416, USA

[6.] Kavli Institute at Cornell for Nanoscale Science, Cornell University, Ithaca, NY, 14850, USA

† Corresponding authors: david.a.muller@cornell.edu; lance.li@kaust.edu.sa

**Two-dimensional (2D) materials are among the most promising candidates for next-generation electronics due to their atomic thinness, allowing for flexible transparent electronics and ultimate length scaling[1]. Thus far, atomically-thin p-n junctions[2-8], metal-semiconductor contacts[9-11], and metal-insulator barriers[12-14] have been demonstrated. While 2D materials achieve the thinnest possible devices, precise nanoscale control over the lateral dimensions are also necessary. Here, we report the direct synthesis of sub-nanometer-wide 1D $MoS_2$ channels**

**embedded within WSe$_2$ monolayers, using a dislocation-catalyzed approach. The 1D channels have edges free of misfit dislocations and dangling bonds, forming a coherent interface with the embedding 2D matrix. Periodic dislocation arrays produce 2D superlattices of coherent MoS$_2$ 1D channels in WSe$_2$. Using molecular dynamics simulations, we have identified other combinations of 2D materials where 1D channels can also be formed. The electronic band structure of these 1D channels offer the promise of carrier confinement in a direct-gap material and charge separation needed to access the ultimate length scales necessary for future electronic applications.**

Reducing the lateral scale of atomically thin 2D devices is crucial not only to realize competitive electronic device applications, but also for reaching the length scales needed for quantum confinement. Thus far, the many 2D heterostructure devices rely on the lithographic patterning of one 2D layer followed by the growth of another in the patterned areas[9-12]. While this technique provides spatial control down to below a hundred nanometers or so, the nature of the lithographic patterning creates atomic defects and contamination. Consequently, the atomic junctions in these heterostructures contain electronic defect states, impacting device performance. Recently, the growth of micron sized in-plane epitaxial interfaces between 2D materials has been reported using chemical vapor deposition (CVD) methods[3-8,13,14]. Theory predicts a tunable carrier confinement[15] and formation of 1D electron gas[16] at the abrupt and coherent interfaces in heterostructures of 2D materials that are just a few atoms wide. The atomic-scale heterostructures are usually chosen for computational convenience, but there would be benefits to realizing such narrow physical structures. For example, in contrast to broad in-plane heterostructures which ultimately generate misfit dislocations to release the lattice strain, thin channels can sustain large strains without relaxation and hence access a wider range of electronic band structures. Just as in bulk materials, dislocation formation can be suppressed below a critical film width[17], which scales inversely with the desired strain – several nanometers are typical for mismatch in the family of 2D transition metal dichalcogenides (TMDs). The thin channels always have one dimension below their critical thickness, ensuring stability against dislocation formation at the strained epitaxial interfaces. Eliminating interfacial dislocations, whose cores are 4|8 or 5|7 member ring structures, is key as these are generally expected to introduce undesirable mid-gap states[15,18,19].

Here, we report an approach for fabricating coherent 1D channels within 2D heterostructures (Fig. 1). These channels possess sub-nanometer widths and atomically coherent sidewalls free of misfit dislocations and dangling bonds. We start with a lateral interface between two 2D TMDs, $MoS_2$ and $WSe_2$, whose lattice mismatch provides an array of interfacial misfit dislocations (Fig. 1b and Supplementary Fig. 1). We introduce growth precursors that provide a high chemical potential for the channel material. The higher reactivity in the core of the misfit dislocations allows the channel atoms (Mo and S) to be inserted into the dislocation core, thus pushing the dislocations away from the original interface, forming 1D $MoS_2$ channels in a trail behind the advancing core (Fig. 1c and Supplementary Fig. 2). The dislocation-catalyzed growth is essentially the flat analog of the semiconductor nanowires whose growth from seeded catalysts has played an important role in semiconductor nanoscience. (See Methods for details on sample preparation and synthesis)

Atomic resolution annular dark field scanning transmission electron microscopy (ADF-STEM) imaging shows that the epitaxial interface between the body of the channel and the host matrix is coherently connected (Fig. 2a and 2b). Meanwhile, a pentagon-heptagon (5|7) dislocation (heptagon pointing up) is found at the terminus of all 1D channels (Fig. 2b). The difference in the atomic number between Mo and W provides high contrast between the $WSe_2$ template and the newly grown $MoS_2$ channels in the ADF-STEM images. Supplementary Fig. 3 shows color coded ADF-STEM images that make the lighter Mo and S atoms more visible. (See Methods for ADF-STEM details)

The as-grown heterostructures of the TMDs must contain strain, due to the bond mismatch to create an epitaxial interface. Applying a geometric phase analysis (GPA)[20] to the ADF-STEM image in Fig. 2a, we are able to elucidate the strain distribution in and around this 1D channel in its 2D matrix, as plotted in Fig. 2c-f (see Supplementary Fig. 4 and Methods for more details). For GPA, the $WSe_2$ lattice parameter was chosen as the reference or zero strain (-.036 would correspond to relaxed $MoS_2$ sheets, consistent with the 3.6% lattice difference measured from the electron diffraction of the $MoS_2$ and $WSe_2$ layers in Supplementary Fig. 1a). Along the x-axis, there is significant difference in the strain map between the 2D $WSe_2$ and 1D $MoS_2$ channels, arising mainly from the lattice mismatch (Fig. 2c). In contrast, the y-axis strain map reveals that $MoS_2$ channels have an identical lattice spacing with the host $WSe_2$ (Fig. 2d), indicating a high uniaxial tensile strain along the y-direction (See Supplementary Fig. 5 for detailed strain analysis on a single channel). Therefore, the newly synthesized 1D channel maintains coherency with the $WSe_2$ matrix and is strain accommodated, which effectively avoids the generation of misfit dislocations along the channel. The shear map and rotation map (Fig. 2e and f) display the position and orientation of the dislocations as dipole fields, confirming all dislocations have the same orientation and migrate upwards (i.e. away from the original hetero-interface).

Growth of the 1D channels is not limited to the interfacial misfit dislocations at the heterostructure interface of the two 2D materials. They can also be generated from intrinsic 5|7 dislocations implanted within the $WSe_2$ film. The ADF-STEM image and corresponding $\varepsilon_{xx}$ strain map (Fig. 2g) of a $MoS_2$ 1D channel show that it was formed from an intrinsic catalyst dislocation migrating in the direction of the heptagon (additional information provided in Supplementary Fig. 6). The isolated 1D channel is 70 nm in length and 1.5 nm in width, surrounded by monolayer $WSe_2$ on all sides, showing a high-aspect-ratio of about 47:1 (length to width).

To understand the catalytic role of 5|7 dislocations, we utilized a reactive force field with newly developed parameters based on density function theory (DFT) calculations and an accelerated molecular dynamics (MD) simulation (details in Supplementary Discussions 1 - 3). Our method captures the dynamics of the chemical bonds breaking and reforming, which is difficult to model using conventional non-reactive MD simulations. We found that unlike other hexagonal rings, the dislocation core allows the precursor atoms to be inserted, which acts as the driving force for the dislocation-catalyzed growth. The precursors first open the catalyst 5|7 dislocation, which admits the Mo insertion (Fig. 3a). This step makes S atoms insufficient to finish the dislocation migration, leaving unsaturated dangling bonds in the system, and hence

the lattice around the dislocation core reconstructs to find more energetically favorable structures. Afterwards, as more S atoms are absorbed from the environment, the lattice relaxes and forms the next dislocation (Fig. 3b). While repeating these two key steps, the previously occupied W and Se atoms near the dislocation core have a certain probability to leave the 2D sheet, and those sites are replaced by the Mo and S precursors during the reconstruction and relaxation. Altogether, the entire process eventually leaves a narrow $MoS_2$ trail behind it (see Supplementary Discussion 4 and Supplementary movies #1-#8 for more details). The circles in Fig. 3c indicate the additional Mo and S atoms placed at the dislocation during each migration step of the catalyst. The additional Mo and S atoms contribute to a 1.4% compressive strain in the x-direction within the channels (Supplementary Fig. 5c).

Due to the crystal geometry of the hexagonal lattice, the migration of the dislocation has two choices: 30º to the right or left (blue or red arrows in Fig. 3c), where the reference lattice orientation is shown as the gray hexagons. MD simulation shows the lateral strain field provides a local restoring force that guides dislocations back towards a straight line along the interface normal, as shown in Supplementary Fig. 7. Thus, the dislocation zigzags about a straight line perpendicular to the macroscopic interface and is ultimately oriented in the heptagon direction, as shown in Fig. 3d.

The dislocation movement out of its slip plane (*climb*[21]) also occurs in a 3D epitaxial interface, due to the diffusion of vacancies or interstitial atoms. In the bulk, this typically does not produce any major effects. In contrast, misfit dislocations in 2D materials can directly take (release) atoms from (to) the environment, suggesting persistent *climbs* that can be used to pattern 1D channels by controlling the precursors and growth time. Statistically, 76% of dislocations tend to migrate and form 1D channels under our optimized growth conditions. Dislocations that did not move tended to have complicated local interface geometries (Supplementary Fig. 1). After measuring 150 1D channels, we achieved an averaged distance between neighboring 1D channels of 10.9 ($\pm$ 0.9) nm, indicating a density of 92 ($\pm$ 8) 1D channels per micron along the interface between $MoS_2$ and $WSe_2$ (Supplementary Fig. 8a). A histogram of channel length is shown in Supplementary Fig. 8b, of which the longest channels reached 80 nm. The length of the 1D channels was strongly correlated with the width of the $MoS_2$ layer around the $WSe_2$ triangles, which is mainly determined by the precursor ratio (S:Mo) and the growth time (Supplementary Fig. 9), suggesting these are two key underlying control parameters. However, there is a limit to how long the $MoS_2$ channels can be grown. As the surrounding $MoS_2$ layer continues to grow, the channel growth ultimately becomes unstable – the 1D channels have possibility to branch repeatedly and recursively, leading to tree-like structures that eventually consume the host material (Supplementary Fig. 10 and 11).

Despite a variety of lengths, more than 90% of the 1D channels have widths that are less than 2 nm, confirming the high accuracy of the dislocation-guided patterning process (Supplementary Fig. 8c). For these ultra-narrow $MoS_2$ channels in $WSe_2$, DFT calculations (Supplementary discussion 5) show a type II band alignment useful for highly-localized carrier confinement and charge separation (Supplementary Fig. 12). Moreover, the strained 1D $MoS_2$ shows a direct band gap, distinct from the indirect bandgap found in uniaxially strained 2D $MoS_2$ thin films[22]. In addition, the 1D channel sidewalls should be free of undesirable mid-gap states that occur due to dislocations and dangling bonds. Both are desirable properties for ultra-small monolayer electronic and optoelectronic applications.

The nature of the 1D growth can be used to create lateral 1D superlattices in 2D materials starting from a periodic dislocation chain, as illustrated in Fig. 4a. The most common structures with periodic dislocations are the grain boundaries of 2D materials[18,19,23,24]. At a typical low-angle $WSe_2$ grain boundary, where two grains with small rotation angles connect laterally to form a classic low-angle tilt boundary, the periodic arrays of 5|7 dislocation cores line up with a spacing ~b/θ. Here, b is the Burger's vector and θ is the tilt angle between the two grains, suggesting grain boundary tilt angle can be used to control the 1D channel spacing. In theory, the dislocations are most stable when they lie vertically above one another with equal spacing[21]. To attain the lowest energy over large scales, the dislocations at the originally curved grain boundary (blue dashed line in Fig. 4b) migrate with an angle of 30º to the left (or to the right) of the heptagon direction (as indicated in Fig. 3c) to form a straight grain boundary. This is also observed in the GPA in Supplementary Fig. 13.

The magnified ADF-STEM image (Fig. 4c) shows a region where all catalyst dislocations migrate 30º to the left forming ~1 nm nanowire arrays with sub-nanometer spacing. Fig. 4d to 4g present the strain maps of Fig. 4c, indicating that dislocations keep their periodicity and orientations after the translation, and the right-side lattice orientation is inherited. We note that in Fig. 4c, short branches appear also on the right side of the original grain boundary, but they have no dislocations at the ends. This can be understood as arising from individual dislocation wandering before they are propelled towards the left by other dislocations, suggesting a strong collective interaction between dislocations that can be used to control the patterning of 1D superlattices. This 1D superlattice formation is commonly observed at low-angle tilt grain boundaries lower than 10º (Supplementary Fig. 14 shows another example).

Our strategy to produce dislocation-free 1D channels suggests a general set of search criteria for other 2D materials. First, candidate materials need a source of dislocations such as low-angle grain boundary or lattice mismatched hetero-interface. Secondly, while the dislocations allow for an easier insertion and exchange of atoms, the substitutions need to be energetically favorable (e.g. S for Se). We used MD simulations to identify another two candidate systems for 1D channels formation: 1D $WS_2$ in $WSe_2$ (a different 1D channel material) and 1D $MoS_2$ in $MoSe_2$ (a different matrix material), both of which are lattice mismatched (see Supplementary Fig. 15 for simulation details). However, combinations of materials that have little lattice mismatch, such as $MoS_2$ and $WS_2$, will not form 1D channels due to the lack of an initial source of catalyst dislocations (see Supplementary Fig. 16 for an experimental example). The lattice mismatch and displacement criteria allows us to predict candidate systems for 1D channel formation, and also provides a way to engineer the strain in the 1D channels by changing the 2D hosts.

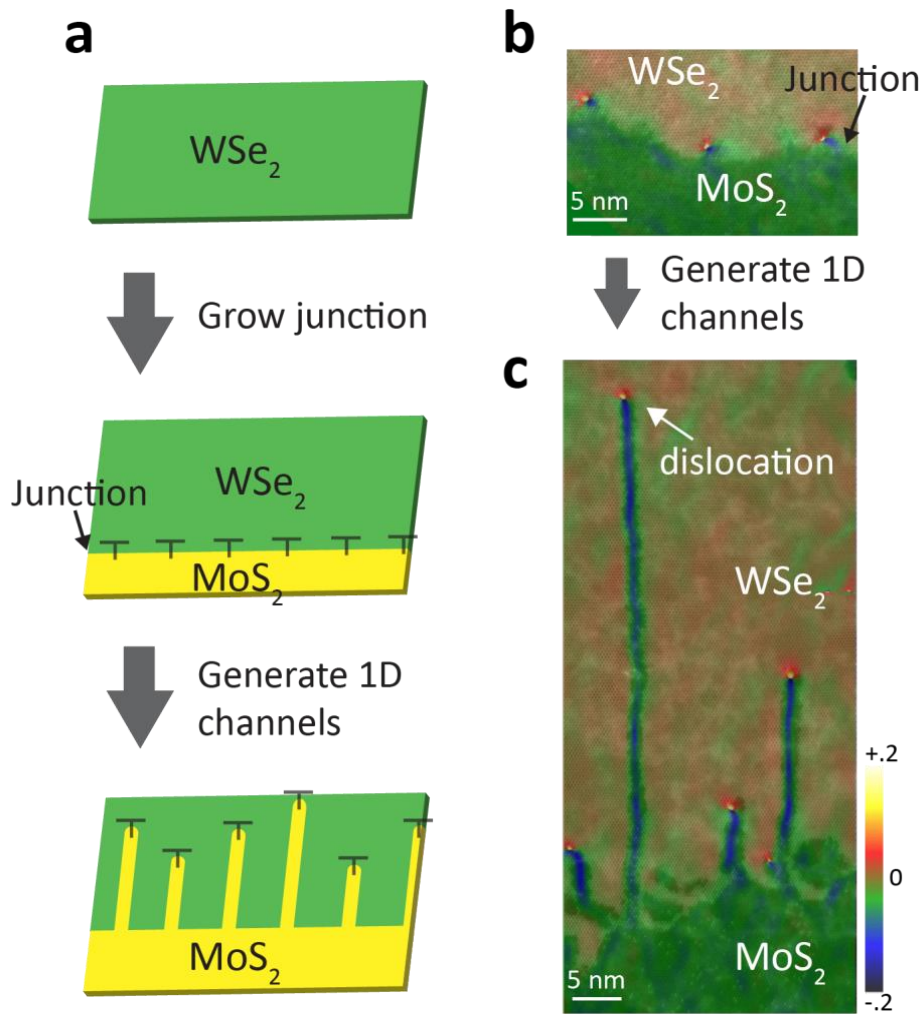

**Figure 1 | Formation of 1D channels. a,** Schematic of the patterning process guided by misfit dislocations (marked as "T") at the MoS$_2$-WSe$_2$ lateral heterojunction. **b** and **c**, atomic resolution ADF-STEM images overlaid with its $\varepsilon_{xx}$ strain maps (see Fig. 2 for more details) identifying the periodic dislocations at the interface of MoS$_2$ and WSe$_2$ (**b**) and the 1D channels created by chemically-driven migration of the interfacial dislocations as additional S and Mo atoms are added (**c**). Strain maps refer to the WSe$_2$ lattice.

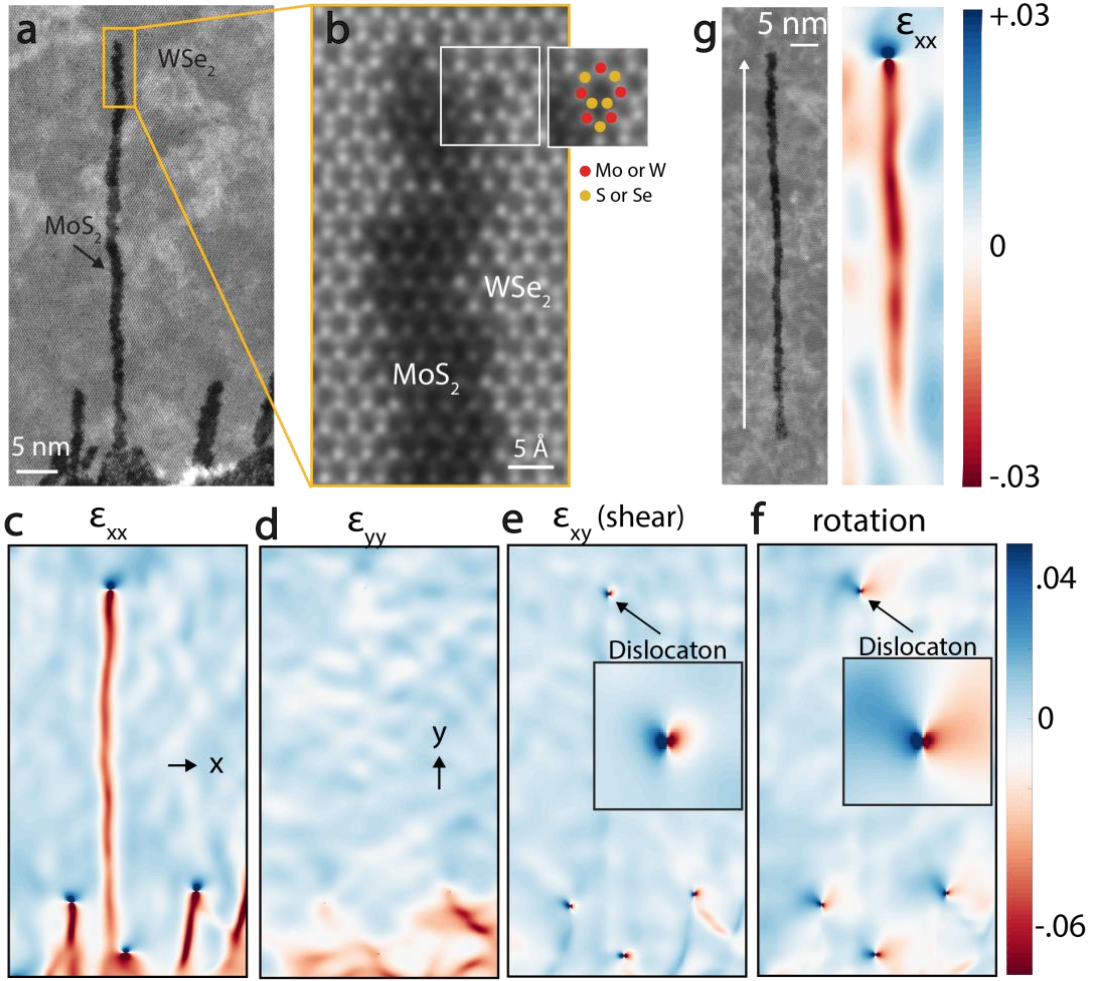

**Figure 2 | Strain maps of the 1D channels**. **a** and **b**, ADF-STEM image of $MoS_2$ 1D channels embedded within $WSe_2$. The channel ends with the 5|7 dislocation (white box in **b**). The same section is shown to the right with the atoms labeled. **c** and **d,** Geometric phase analysis (GPA) of the 1D $MoS_2$ in **a** with uniaxial strain components $\varepsilon_{xx}$ (**c**) and $\varepsilon_{yy}$ (**d**). All the strain is in reference to the $WSe_2$ lattice. The $\varepsilon_{xx}$ clearly distinguishes the two lattices mainly due to the lattice mismatch, while the $\varepsilon_{yy}$ indicates a high uniaxial tensile strain in the 1D $MoS_2$ which is lattice mismatched from the $WSe_2$. **e** and **f** display the shear strain and the rotation map (in radians) indicating the position and orientation of the dislocations. **g,** ADF-STEM image and its $\varepsilon_{xx}$ strain map of a $MoS_2$ 1D channel formed from an intrinsic 5|7 dislocation in

WSe$_2$, which matches the results found in channels arising from the heterojunction interface.

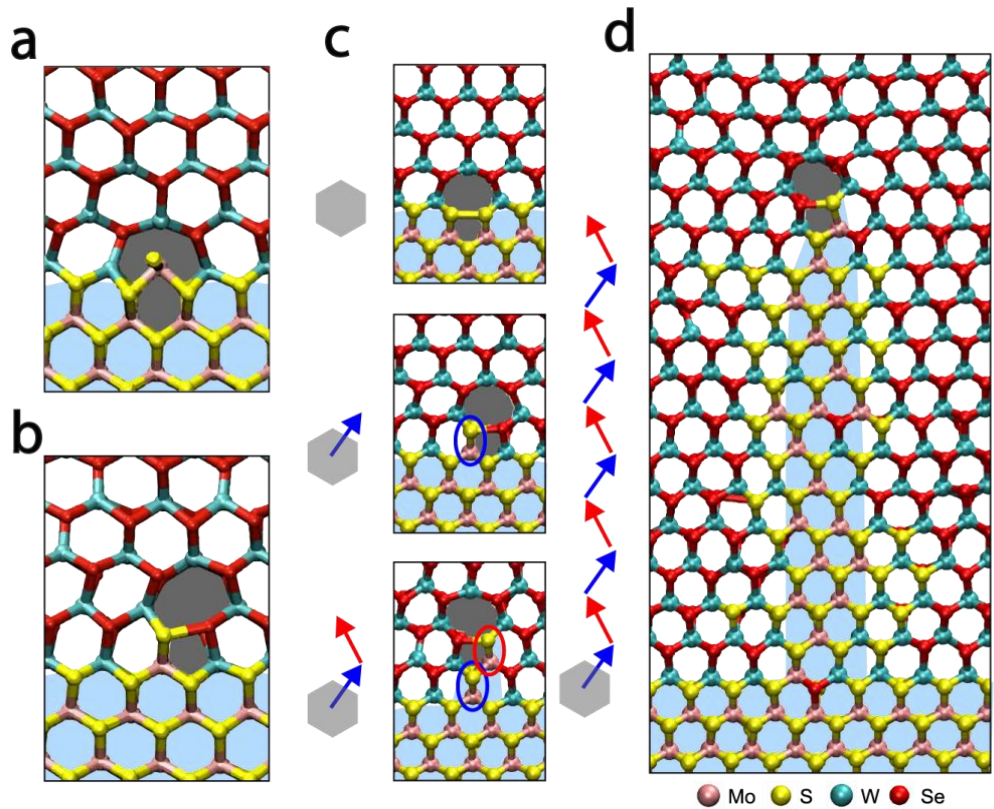

**Figure 3 | Molecular dynamics (MD) simulation of the 1D channel formation. a, b,** MD simulation of the process of Mo inserting into pentagon ring of 5|7 dislocation (**a**) and the formation of the next 5|7 dislocation (**b**). **c**, MD simulation of each step for the patterning process. The 5|7 dislocation can migrate 30º to the right (blue arrow) or 30º to the left (red arrow). **d**, The iteration of adding excess Mo and S atoms forms the 1D MoS$_2$ channel in WSe$_2$, unveiling the chemically driven mechanism for the formation of the 1D channels.

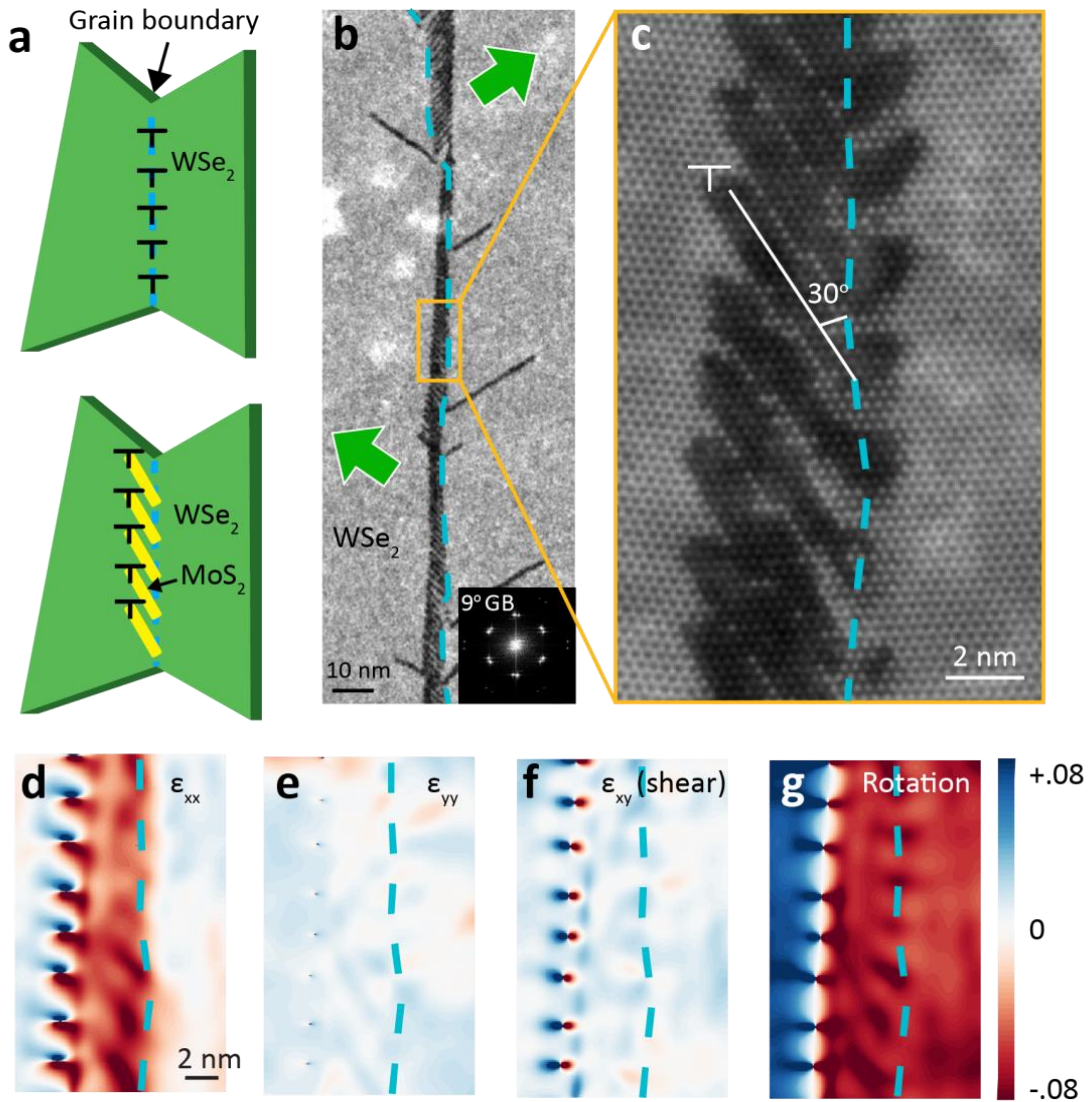

**Figure 4 | Generation of a super lattice at a grain boundary. a,** Schematic of the superlattice formation where the top (bottom) panel depicts the grain boundary before (after) the patterning process. **b,** ADF-STEM image of a superlattice grown from the periodic dislocations at the WSe$_2$ grain boundary with 9° rotation (2 nm spacing between dislocations). All blue dashed lines indicate the position of the original curved grain boundary. The dislocations migrate in different directions (indicated by the green arrows), thus forming a shifted but straight grain boundary. **c,** Magnified ADF-STEM image with one of the identical dislocations marked by a "T". **d-g,** GPA of (**c**) showing that dislocations preserve their periodicity and orientations during the migration.

**Acknowledgements** The authors acknowledge discussions with Mervin Zhao, Lei Wang, Chen Zhen, Megan Holtz, Heung-Sik Kim, Cheng Gong, Ting Cao, Mariena S. Ramos, Lena F. Kourkoutis, Benjamin Savitzky, Mengnan Zhao, Cheol-Joo Kim, Kibum Kang, Jiwoong Park, Debdeep Jena, James Sethna. This work made use of the electron microscopy facility of the Cornell Center for Materials Research (CCMR) with support from the National Science Foundation (NSF) Materials Research Science and Engineering Centers (MRSEC) program (DMR-1120296) and NSF Major Research Instrumentation Program (DMR-1429155). Y.H. and D.A.M. are supported by NSF Grant (DMR-1719875) and DOD-MURI (Grant No. FA9550-16-1-0031). G.S.J, Z.Q, and M.J.B acknowledge the support by the Office of Naval Research (Grant No. N00014-16-1-233) and DOD-MURI (Grant No. FA9550-15-1-0514). We acknowledge support for supercomputing resources from the Supercomputing Center/KISTI (KSC-2017-C2-0013). M.L. and L.L. thank the support from King Abdullah University of Science and Technology (KAUST) and Academia Sinica.

**Author Contributions** Y.H., M.L., and G.-S. J. contributed equally to this work. Y.H. conceived the project. Electron microscopy and data analysis were carried out by Y.H. under the supervision of D.A.M. with the help from M.A.M. Sample growth was done by M.L., under the supervision of L.L.. The molecular dynamics simulations and density function theory calculations were conducted by G.S.J. and Z.Q., under the supervision of M.J.B.

**METHODS**

**Sample growth.** First, the WSe2 monolayers were grown on sapphire substrates using the chemical vapor deposition (CVD) method, with the WO3 and Se powders as the precursors carried by Ar (90 sccm) and H2 (6 sccm) gases. The furnace was heated to 925 oC at a pressure of 15 Torr for 15 min to grow WSe2 monolayers, which contain sharp edges and grain boundaries. After cooling, the sample was placed in another furnace for the second step growth of MoS2 monolayers, forming abrupt epitaxial junctions with misfit dislocations. During the second step, MoO3 (at 760 oC) and S (at 190 oC) powders were used as the precursors carried by Ar gas flowing at 70 sccm at a pressure of 40 Torr and a temperature of 760 oC for 10 minutes. However, in the second step, since the dislocations along the abrupt junctions and grain boundaries had already been exposed to the Mo and S precursors at such a high temperature, the patterning process had already begun and the 1D MoS2 had been formed.

**Transfer to TEM grids.** The sample was coated with PMMA A4 to support the film during the transfer process. To detach the film from the sapphire substrate, the sample was placed in HF solution (HF:$H_2$O 1:3) for 15 minutes. After rinsing with DI water several times, the sample was blow-dried and dipped into water – using the surface tension – to release the film from the substrate. With the film floating on the surface of the water, we applied a QUANTIFOIL holey carbon TEM grid to scoop the film. Afterwards, the TEM grid with samples was baked in vacuum (1 x $10^{-7}$ torr) at 350 oC for 5 hours to remove the PMMA. Baking in vacuum is essential because the dislocations will degrade and form holes if baked in air.

**ADF-STEM.** ADF-STEM imaging was conducted using an aberration-corrected FEI TITAN operated at 120 kV with a ~15 pA probe current. The acquisition time per pixel was less than 8 milliseconds, but multiple images (10~20) were acquired and cross-correlated afterwards to improve the signal-noise-ratio and reduce the scan noise introduced by the sample drift. Despite the electron beam energy above the knock-on damage threshold for $WSe_2$ and $MoS_2$, the low dose per image, in fact, avoided significant damage from ionization. A 30 mrad convergence angle and a ~40 mrad inner collection angle were used for all ADF-STEM images, whose contrast is proportional to $Z^{\gamma}$, where Z is the atomic number and $1.3 < \gamma < 2$. Therefore, the W, Se, Mo and S atoms can be distinguished easily by this Z-contrast imaging technique.

**Geometric phase analysis.** Geometric phase analysis (GPA) is a method for measuring and mapping strain fields from high-resolution electron microscope images[25]. It describes how the spatial frequency components (lattice fringes) of the image vary across the image field of view. Here in this work we applied GPA to atomic resolution ADF-STEM images of $MoS_2$ 1D channels embedded within $WSe_2$ monolayers. We used the GPA plugin[26] developed for Digital Micrograph, and the detailed process was described below and in Supplementary Fig. 4.

1) Fourier transform the lattice image: We firstly Fourier transform the atomic resolution images (Supplementary Fig. 4a) to the power spectrum (Supplementary Fig. 4b). In the power spectrum, the strong Bragg-reflections are related to the unit cell of the crystalline structure of the material. A perfect crystal lattice gives rise to sharply peaked frequency components, while the broadening of the Bragg spots is due to the local lattice distortion in the material.

2) Place masks: Instead of using a mask covering the entire first Brillouin zone, practically we placed circular Gaussian masks on two non-colinear reciprocal lattice vectors g₁ and g₂, as shown in Supplementary Fig. 4b in red and blue circles on the power spectrum. The size of the masks is smaller than the Brillouin zone. The resolution and smoothing setup in Supplementary Fig. 4k define the size and smoothing of the masks, which help to reduce noise and smooth the resulting images.

3) Calculate the phase image: We convolved each region around reciprocal vector g₁ and g₂ with the masks. Afterwards, we performed an inversed Fourier transform to create a complex image that the phase image was calculated from.

$$P_g(\mathbf{r}) = \text{Phase}[H'_g(\mathbf{r})] - 2\pi \mathbf{g} \cdot \mathbf{r}. \tag{1}$$

where $H_g'(\mathbf{r})$ is the complex image from the inversed Fourier transform, and **g** is the reciprocal lattice vector where the mask was placed. The phase images corresponding to reciprocal lattice vector **g₁** and **g₂** were plotted in Supplementary Fig. 4c and 4d after a renormalization between ±π.

4) Determine the displacement field: In the presence of a displacement field **u**, the maximum of the fringes **r** is displaced by **u**, and becomes **r-u**. In this case, we can write the intensity of Bragg filtered images that were produced by the Gaussian mask at **g**:

$$B_g(\mathbf{r}) = 2A_g \cos[2\pi \mathbf{g} \cdot \mathbf{r} - 2\pi \mathbf{g} \cdot \mathbf{u} + P_g]. \tag{2}$$

where the $A_g$ is the amplitude, and $P_g$ is the arbitrary constant phase that can be ignored. From Eq. (2), we note that the middle term is a phase term that depends on the lattice displacement field **u**: $P_g(\mathbf{r}) = -2\pi \mathbf{g} \cdot \mathbf{u}$, which can be calculated by taking inverse of **g**:

$$\mathbf{u}(\mathbf{r}) = -\frac{1}{2\pi}\left[P_{g1}(\mathbf{r})\mathbf{a_1} + P_{g2}(\mathbf{r})\mathbf{a_2}\right]. \tag{3}$$

where $\mathbf{a_1}$ and $\mathbf{a_2}$ are the inverse of $\mathbf{g_1}$ and $\mathbf{g_2}$.

5) Determine the strain and rotation fields: The local distortion of the lattice can be calculated from the gradient of the displacement field and defined as a 2 by 2 matrix:

$$e = \begin{pmatrix} e_{xx} & e_{xy} \\ e_{yx} & e_{yy} \end{pmatrix} = \begin{pmatrix} \frac{\partial u_x}{\partial x} & \frac{\partial u_x}{\partial y} \\ \frac{\partial u_y}{\partial x} & \frac{\partial u_y}{\partial y} \end{pmatrix}. \tag{4}$$

The strain is given by the symmetric term $\varepsilon = \frac{1}{2}[e + e^T]$ and the rigid rotation is described by the anti-symmetric term $\omega = \frac{1}{2}[e - e^T]$. In this paper, the uniaxial strain can be calculated using $\varepsilon_{xx} = e_{xx}$, and $\varepsilon_{yy} = e_{yy}$, as shown in Supplementary Fig. 4g and 4h respectively. The shear strain field map is shown in Supplementary Fig. 4i calculated using $\varepsilon_{xy} = (e_{xy} + e_{yx})/2$. The rotation map displayed in Supplementary Fig. 4j is calculated by $\varepsilon_{rot} = (e_{xy} - e_{yx})/2$.

In Supplementary Fig. 4k, we cropped the GPA plugin[2] control panel. In this software, 'a*' displays the length of reciprocal lattice vector $\mathbf{g_1}$ in 1/nm unit and 'b*' shows that of $\mathbf{g_2}$. The local $|\mathbf{g_1}|$ and $|\mathbf{g_2}|$ were mapped in Supplementary Fig. 4e and 4f. Gamma represents the angle between $\mathbf{g_1}$ and $\mathbf{g_2}$ in degrees, while theta displays the angle between $\mathbf{g_2}$ and the horizontal axis (white arrow in Supplementary Fig. 4b). The resolution setup defines the size of the Gaussian masks and the smoothing defines the mask edge smoothing. The 'refine G-vectors' button calculate the $\mathbf{g}$ vectors in the reference lattice region we selected, thus refine the center of the masks. Here in this work we select the flat WSe$_2$ region as the reference lattice.

**MD for the growth of a 1D MoS$_2$ channel.** MD simulations in this study were performed via LAMMPS MD package[27] using the Reactive Empirical Bond Order (REBO) force field[28-30] to model the interactions among Mo, W, S, and Se atoms. We utilized the optimized parameters of Mo-S REBO for failure of MoS$_2$ monolayer[31] and developed reliable force field for other atoms with Density Functional Theory (DFT) calculations[31,32]. We built a model composed of 4 nm MoS$_2$ and 6 nm WSe$_2$ in the *y* direction with 7 nm junction regions in the *x* direction along the interface on a simplified substrate. We applied a number of cyclic annealing processes, including relaxation of the structures from a high to a low temperature, and adding/deleting atoms on the basis of Monte Carlo method. All processes are designed to accelerate the evolution of 1D MoS$_2$ nanowire with natural bond forming and structural deforming (See more details in **Supplementary Discussions 1-3**). We utilized VMD tool to visualize the simulations[33].

# Supplementary Information for Sub-Nanometer Channels Embedded in Two-Dimensional Materials

Yimo Han[1]*, Ming-Yang Li[2,3]*, Gang-Seob Jung[4]*, Mark A. Marsalis[5], Zhao Qin[4], Markus J. Buehler[4], Lain-Jong Li[2]†, David A. Muller[1,6]†

[1.] School of Applied & Engineering Physics, Cornell University, Ithaca, NY, 14850, USA

[2.] Physical Science and Engineering Division, King Abdullah University of Science and Technology, Thuwal, 23955-6900, Kingdom of Saudi Arabia

[3.] Research Center for Applied Sciences, Academia Sinica, Taipei, 10617, Taiwan

[4.] Department of Civil and Environmental Engineering, MIT, Cambridge, MA, 02139, USA

[5.] Department of Physics, Texas Tech University, Lubbock, TX, 79416, USA

[6.] Kavli Institute at Cornell for Nanoscale Science, Cornell University, Ithaca, NY, 14850, USA

† Corresponding authors: david.a.muller@cornell.edu; lance.li@kaust.edu.sa

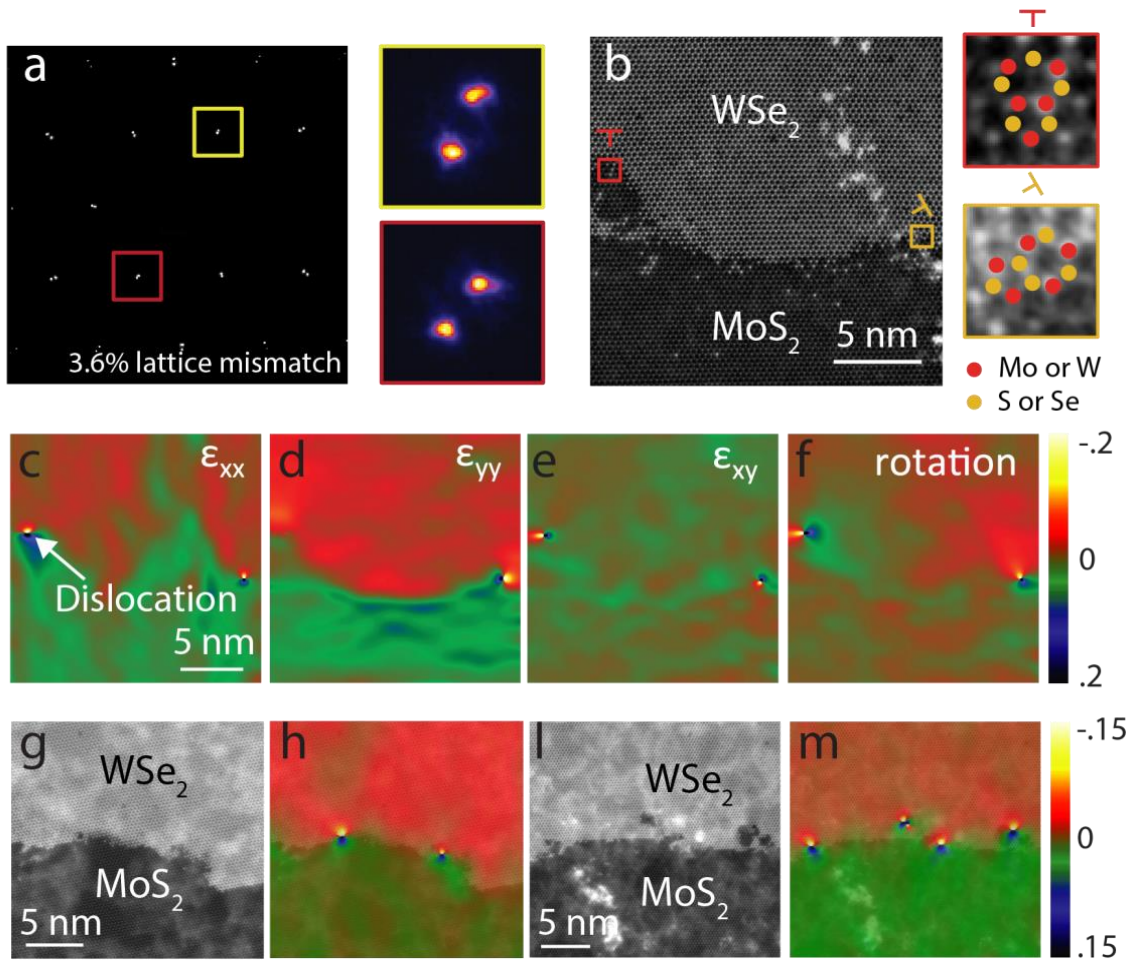

**Supplementary Figure 1 | Dislocations at abrupt MoS$_2$-WSe$_2$ junctions. a,** Diffraction pattern of the abrupt junction from a micron-sized area with magnified diffracted spots on the right, indicating that the two materials are flat and fully relaxed. We fit the peaks to Gaussian and located the centers, showing a 3.6% lattice mismatch between MoS$_2$ and WSe$_2$. **b,** Atomic resolution ADF-STEM image of the abrupt MoS$_2$-WSe$_2$ junction with two misfit dislocations appearing at the interface. The magnified images of the dislocations are shown on the right with the atoms marked. They are pentagon-heptagon pair dislocations with Mo-Mo bonds (red border) or S-S bonds (yellow border). In the formation of 1D MoS$_2$, both types of the 5|7 dislocations behave similarly as catalysts. **c-f,** GPA maps of (**b**), indicating the lattice strain, in addition to the location and orientation of the misfit dislocations. The GPA method is discussed in Methods, Supplementary Fig. 4, and reference 20 in the main manuscript. **g-m,** Additional ADF images (**g, l**) and the overlay with $\varepsilon_{xx}$ strain maps (**h, m**) of abrupt junctions with dislocations at the interface.

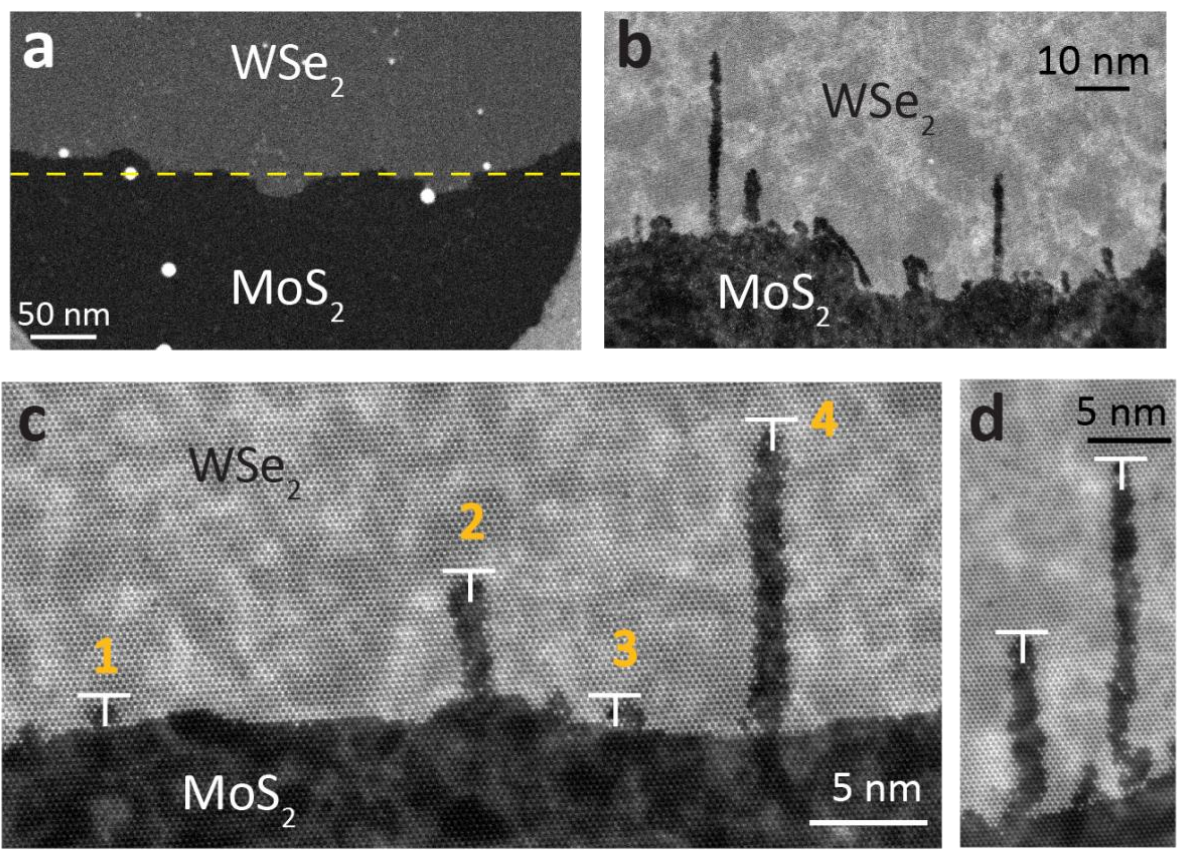

**Supplementary Figure 2 | 1D channels formed by misfit dislocations. a**, Large scale ADF-STEM image of the abrupt junction. **b,** Magnified ADF-STEM image with aligned 1D MoS₂ channels perpendicular to the junction. **c, d**, Atomic resolution ADF-STEM images of the 1D channels with dislocations marked as "T". In (**c**), #1 and #3 form short bumps while #2 and #4 propel 1D channels, indicating a length variation of the 1D channels, which is explained statistically in Supplementary Fig. 8. Approaches to control the length are discussed in Supplementary Fig. 9.

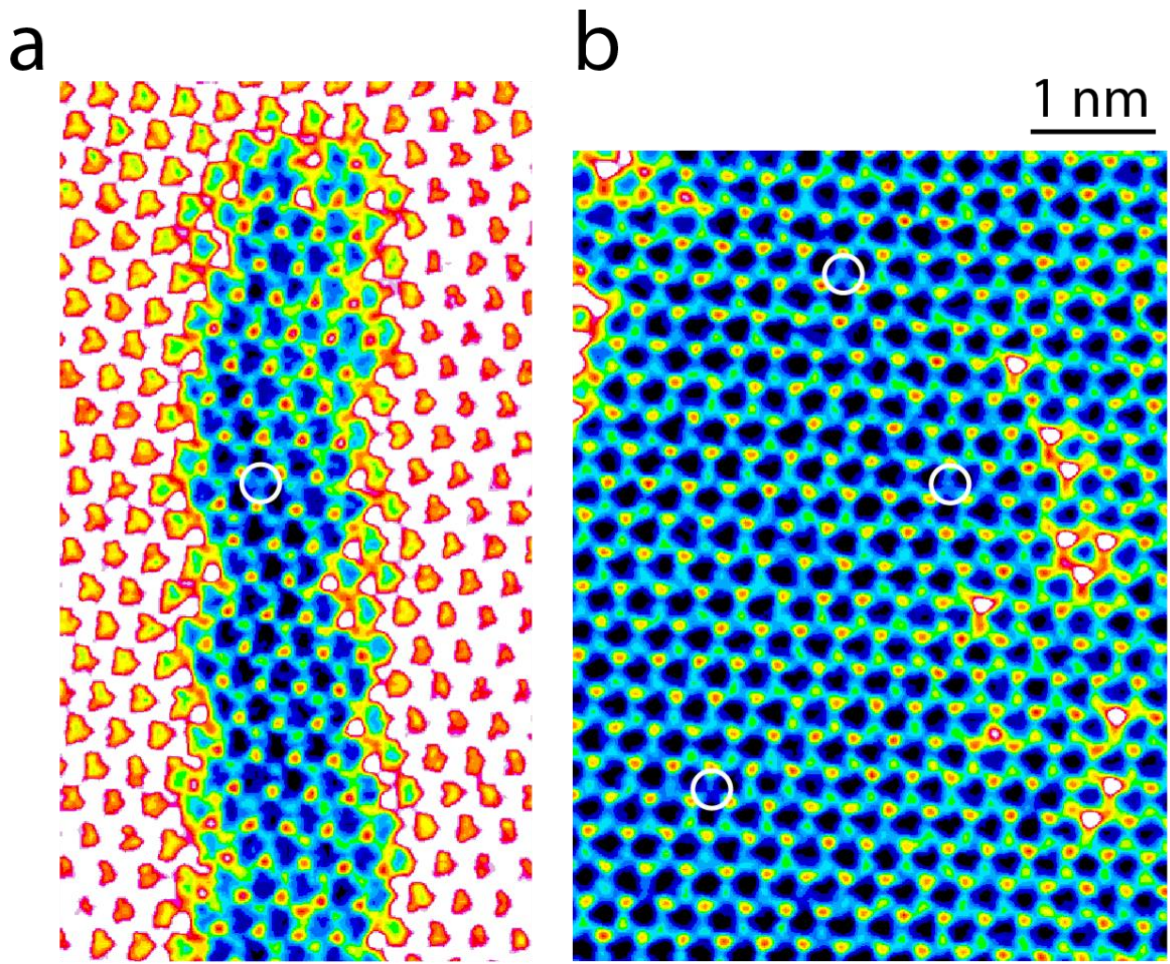

**Supplementary Figure 3 | Color coded ADF-STEM images making lighter Mo and S more visible. a,** Color coded ADF-STEM image of a 1D MoS$_2$ channel with brightness and contrast optimized to make lighter Mo and S atoms more visible. We observed one sulfur vacancy (white circle) in this region, showing a sulfur vacancy density of 0.091 nm$^{-2}$ (1 in 11 nm$^2$). **b,** Color coded ADF-STEM image of a MoS$_2$ sheet with the same color scale as (**a**). We observed three vacancies (white circles), giving a sulfur vacancy density of 0.094 nm$^{-2}$ (3 in 32 nm$^2$). The comparable density of sulfur vacancies is consistent with the rate of our electron beam damage (i.e. we lose a few sulfur atoms every scan).

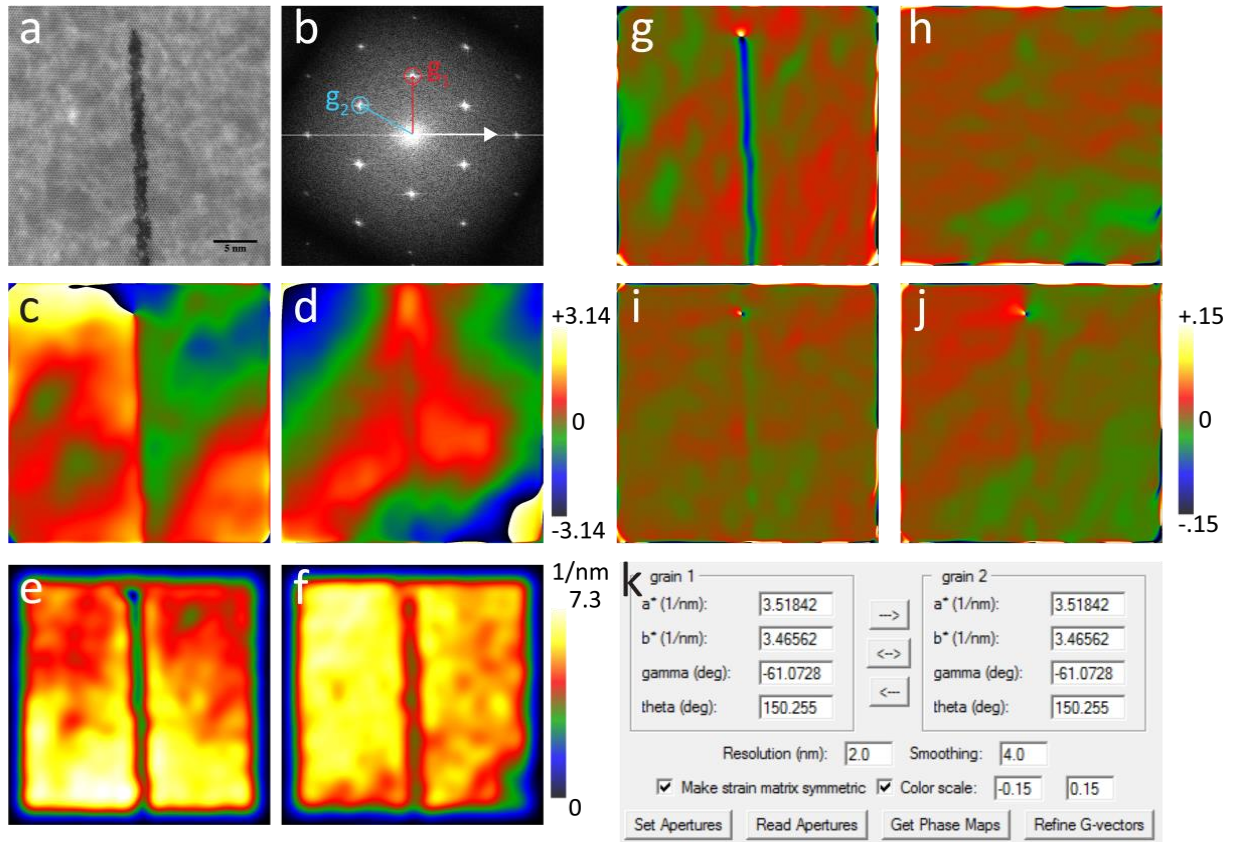

**Supplementary Figure 4 | Geometric phase analysis (GPA). a,** ADF-STEM showing the 1D $MoS_2$ channel embedded within $WSe_2$. **b,** The Fourier transform of **a** with apertures indicated by red and blue circles. **c, d,** The geometric phase images calculated from $g_1$ and $g_2$ in (**b**). **e, f,** The scale of the reciprocal lattice vectors $g_1$ and $g_2$ respectively. **g-j,** The strain maps: $\varepsilon_{xx}$, $\varepsilon_{yy}$, $\varepsilon_{xy}$, and rotation respectively. **k,** The screen capture of the control panel in the GPA plugin. See Methods for more details.

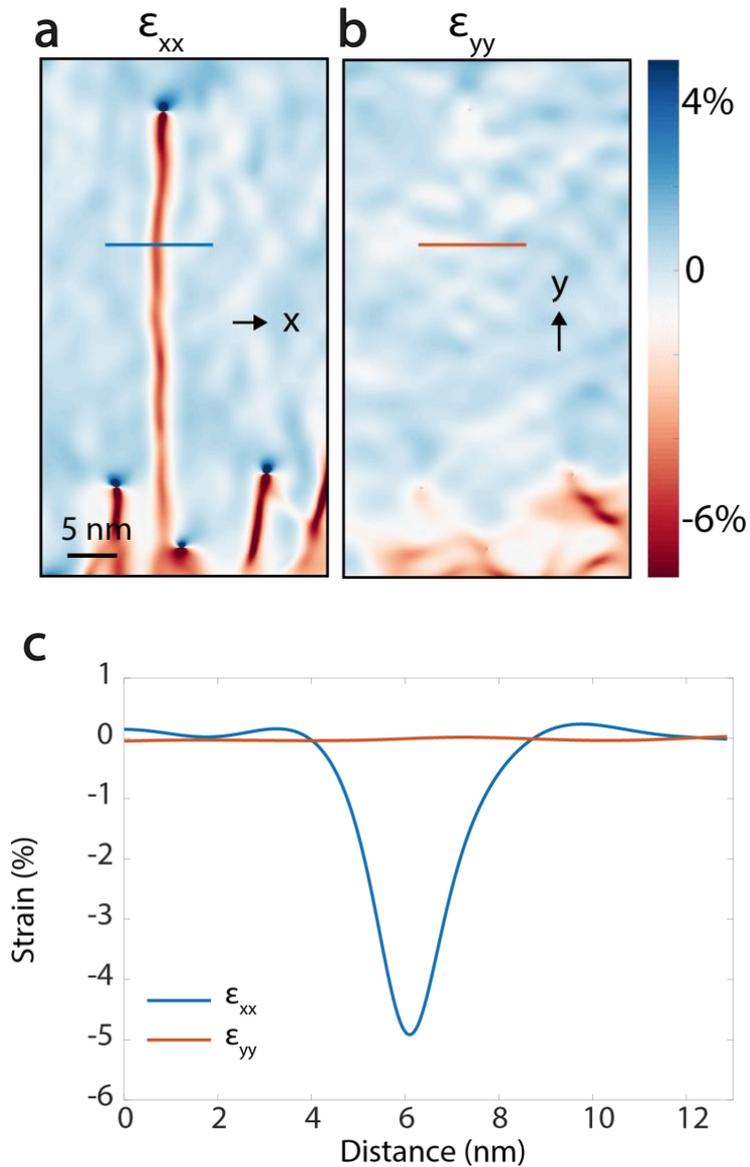

**Supplementary Figure 5 | Uniaxial strain in 1D MoS₂ embedded within WSe₂. a,b,** The uniaxial strain $\varepsilon_{xx}$ and $\varepsilon_{yy}$ maps shown in Fig. 2 in the main manuscript. **c,** The line profile from the blue line region in **a** and red line region in **b** respectively. The minimum (blue line) shows a ~5% lattice difference in the MoS₂ from the surrounding WSe₂, indicating a small compressive strain (~1.4%) along the x direction in the MoS₂ 1D channels, which can be calculated from the 3.6% lattice mismatch between MoS₂ and WSe₂. Along the y direction, the MoS₂ and WSe₂ are lattice matched, showing a 3.6% tensile $\varepsilon_{yy}$ strain in the MoS₂ 1D channel to form the coherent structure.

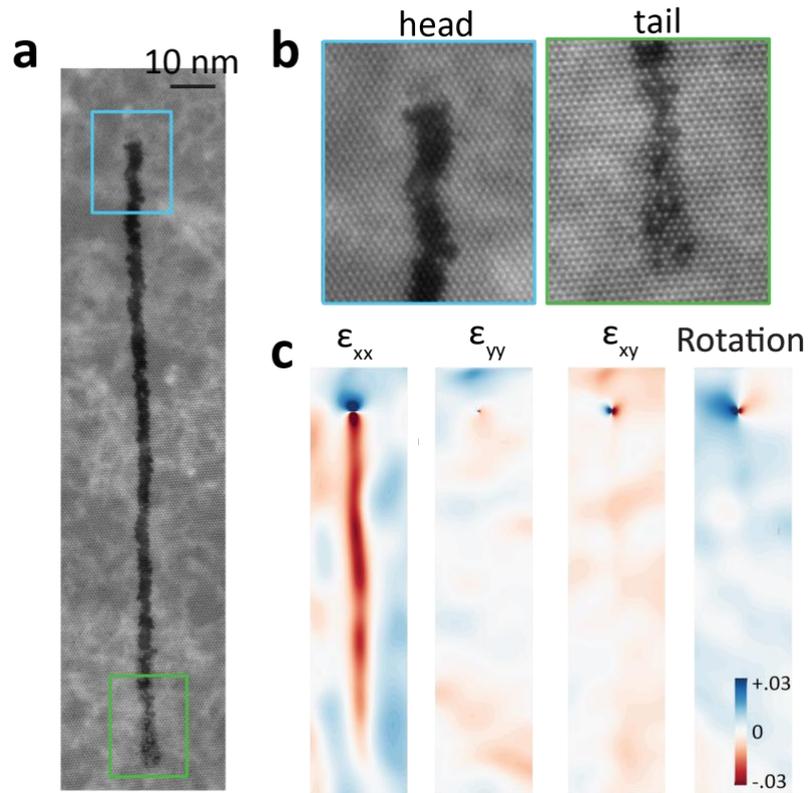

**Supplementary Figure 6 | 1D channels guided by intrinsic dislocations. a,** ADF-STEM image of the 1D channel guided by the intrinsic 5|7 dislocation that is originally embedded within $WSe_2$. **b,** The magnified images of the head (tail) of the 1D channel, showing sharp (alloyed) interfaces. **c**, The GPA maps from (**a**), indicating that the dislocation climbs up, towards its heptagon direction, which is consistent with the 1D channels from interface dislocations (Fig. 1, 2 and Supplementary Fig. 2).

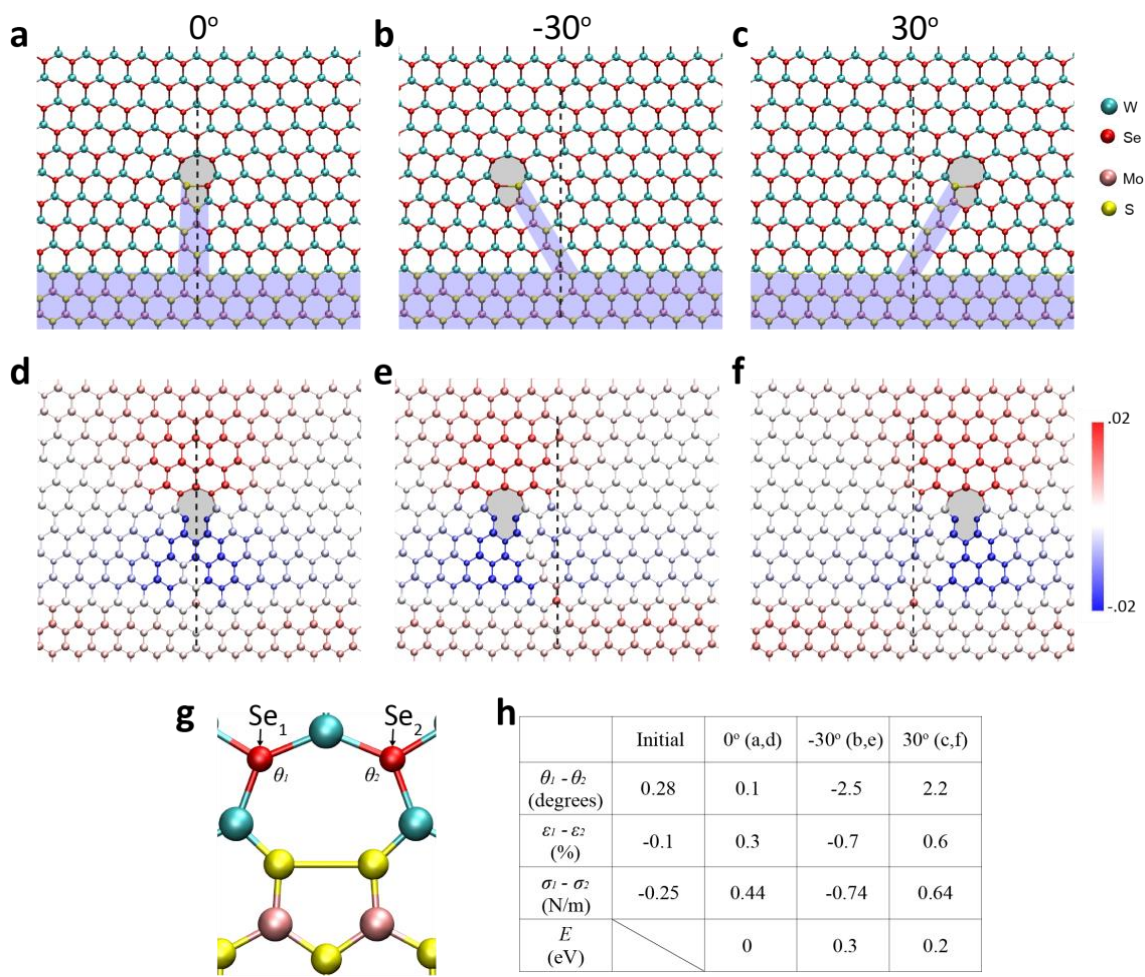

**Supplementary Figure 7 | Strain guides the migration direction. a-c,** Schematic of the 1D channels migrating via a zigzag path perpendicular to the interface **(a)**, path 30º to the left **(b)** and right **(c)**. **d-f,** The corresponding strain maps of **a**, **b**, and **c**. The strain map of 0º 1D channel **(d)** shows a symmetric strain, while **(e)** and **(f)** show clear asymmetric strain. **g,** Schematic of the catalyst dislocation. **h,** Table of the local bond angle ($\theta$), atomic strain ($\varepsilon$), atomic stress ($\sigma$) and total energy (E) differentiations between $Se_1$ and $Se_2$ selenium atoms (**g**) for the three migrating paths shown in **a-c**. The initial one is calculated from the case where the dislocation is at the interface without migrating. The bond angle, strain, and stress differences between the two W-Se bonds reveal a local asymmetry in the dislocation core. This asymmetry affects the breaking of left or right W-Se bonds (i.e. migrating toward left or right) when the precursors insert into the lattice. The comparison of total energy confirms the dislocation prefers to form a zigzag path that is perpendicular to the interface.

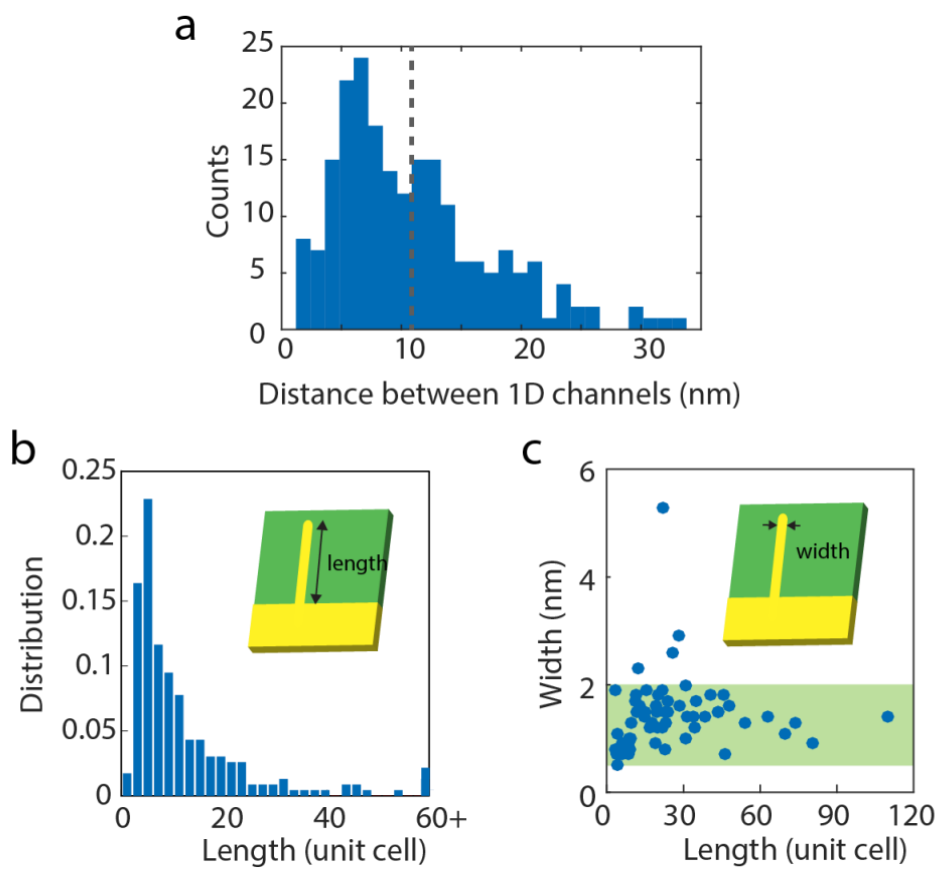

**Supplementary Figure 8 | Statistics of 1D channels. a,** The histogram of the distance between two neighboring 1D channels. The average distance is 10.9 (± 0.9) nm, indicating 92 (±8) 1D channels per micron along the $MoS_2$-$WSe_2$ interface. Meanwhile, the measured mean spacing between dislocations at this interface is 8.3 (± 0.6) nm, indicating a density of 121 (±10) misfit dislocations per micron. This result shows that 76% (± 8%) of the dislocations tend to migrate and form 1D channels. Errors reported are twice the standard error of the mean. **b,** The length distribution of the 1D channels displays an abrupt drop below 2 nm, suggesting that most 1D channels tend to grow once the catalyst dislocations start to migrate. Using our current recipe, the longest channel that has been observed is 80 nm. **c,** The scatter plot of the channel width according to their length shows that more than 90% of the channels have widths less than 2 nm. Overall, we ran more than 10 growth rounds and prepared more than 20 TEM samples. We see 1D channels in all the samples. Our statistics of the length, width, and neighboring distance are from ~150 1D $MoS_2$ channels. The statistics are taken from the growth condition in Supplementary Fig. 9c and 9d, where the $MoS_2$ surrounding the $WSe_2$ triangle is micron-meter-sized.

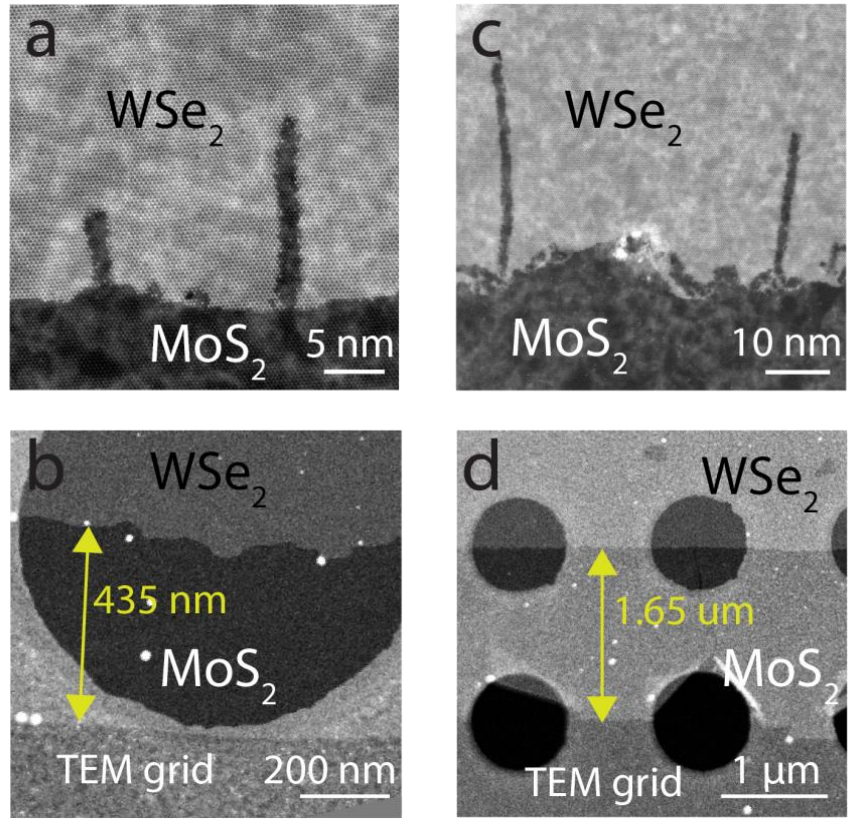

**Supplementary Figure 9 | Controlling the lengths of the 1D channels. a, b,** ADF-STEM images of the 1D channels (**a**) and the overview of the lateral junction (**b**). The MoS$_2$ surrounding the WSe$_2$ triangles is 435 nm (indicated by the yellow arrow) in this sample, while the 1D channel lengths are only several nanometers. **c, d,** ADF-STEM images of the sample with the surrounding MoS$_2$ 1.65 μm wide (yellow arrow). The 1D channels are much longer than those in nanometer-wide MoS$_2$ samples (**b**). In our study, we focused on samples in this regime (**c** and **d**). To conclude, the dislocation tended to form longer channels with wider surrounding MoS$_2$ sheets, which can be controlled by the growth time or the precursor ratio (S:Mo). Slightly increasing the S:Mo ratio helps to grow longer 1D channels. However, S-excess conditions cause the substitution of the WSe$_2$ layers by MoS$_2$ (ref. 6 in main manuscript and Supplementary Fig. 10a and b).

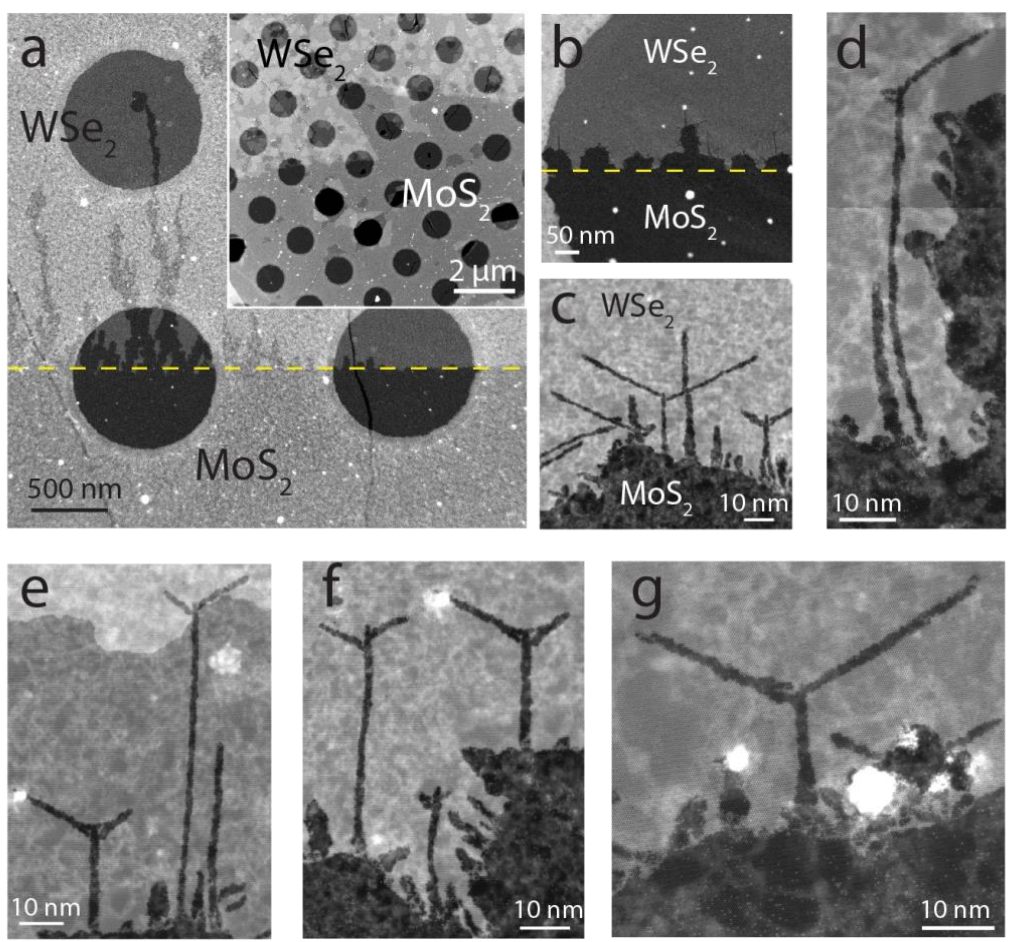

**Supplementary Figure 10 | Branching of 1D channels under sulfur excess growth condition. a,** Large field-of-view ADF-STEM image, where the $MoS_2$ replace the some existing $WSe_2$ and form the dendritic shapes at a $MoS_2$-$WSe_2$ junction (the yellow dashed line) in a sulfur excess growth condition, which was discussed in reference 6 in the main manuscript. The inset shows the outside $MoS_2$ grows ~10 μm or even more from the $WSe_2$ edges, where the $MoS_2$ also nucleate and form a second layer on top of $WSe_2$, which are the small brighter triangles. **b,** Region with the massive replacements can form periodic wave fronts. **c-g,** Catalyst dislocations split into two partial dislocations and form branches. The dislocation splitting and channel branching process are explained in Supplementary Fig. 11.

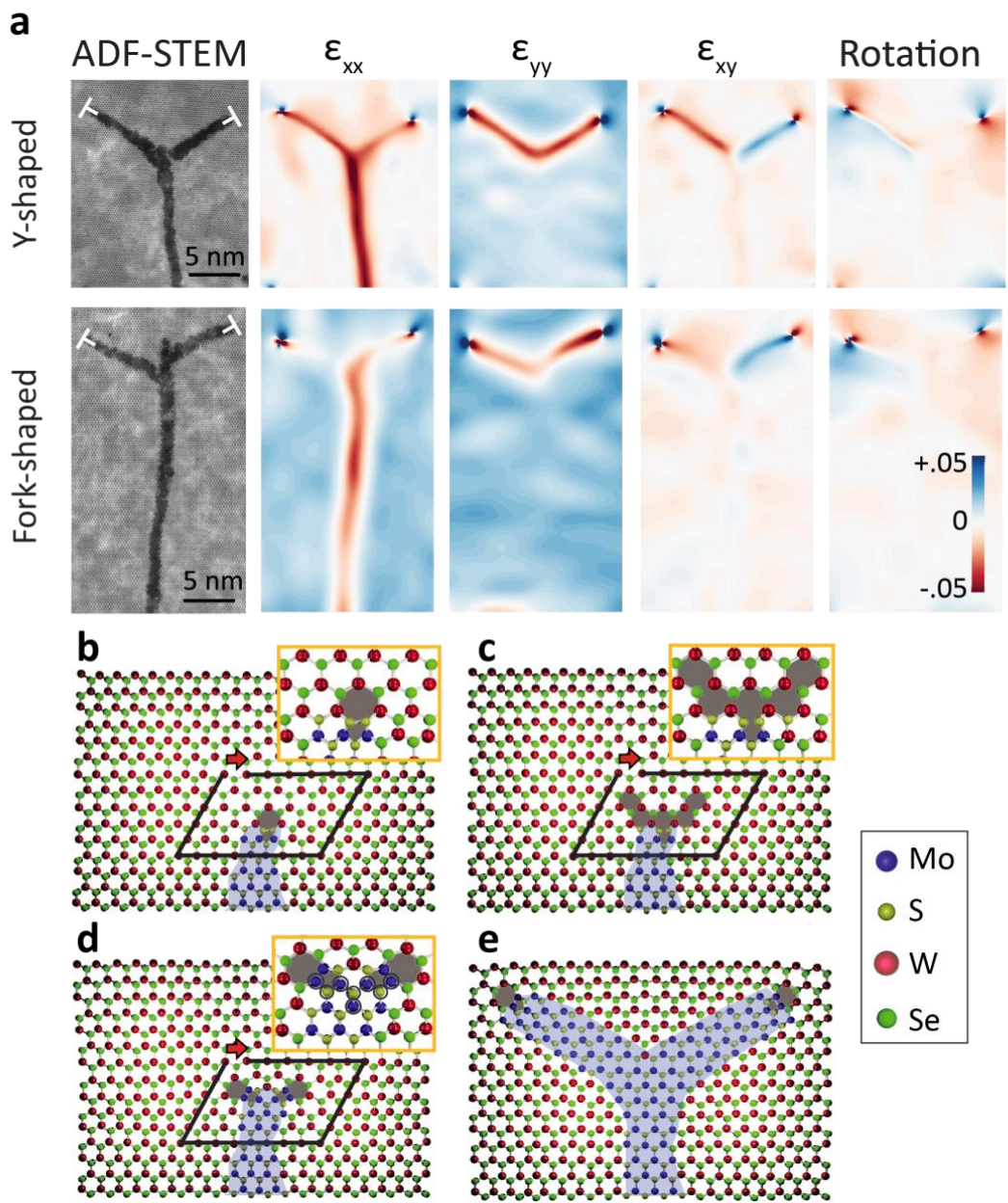

**Supplementary Figure 11 | Dislocation splitting and 1D channel branching. a,** ADF-STEM images and their corresponding strain maps of the Y-shaped and fork-shaped junctions. **b-e,** Atomic model for the dislocation splitting. The red arrows represent the Burger's vector, which is conserved during the splitting process. The black circles indicate the excess Mo and S atoms inserted into the system during the dislocation splitting process.

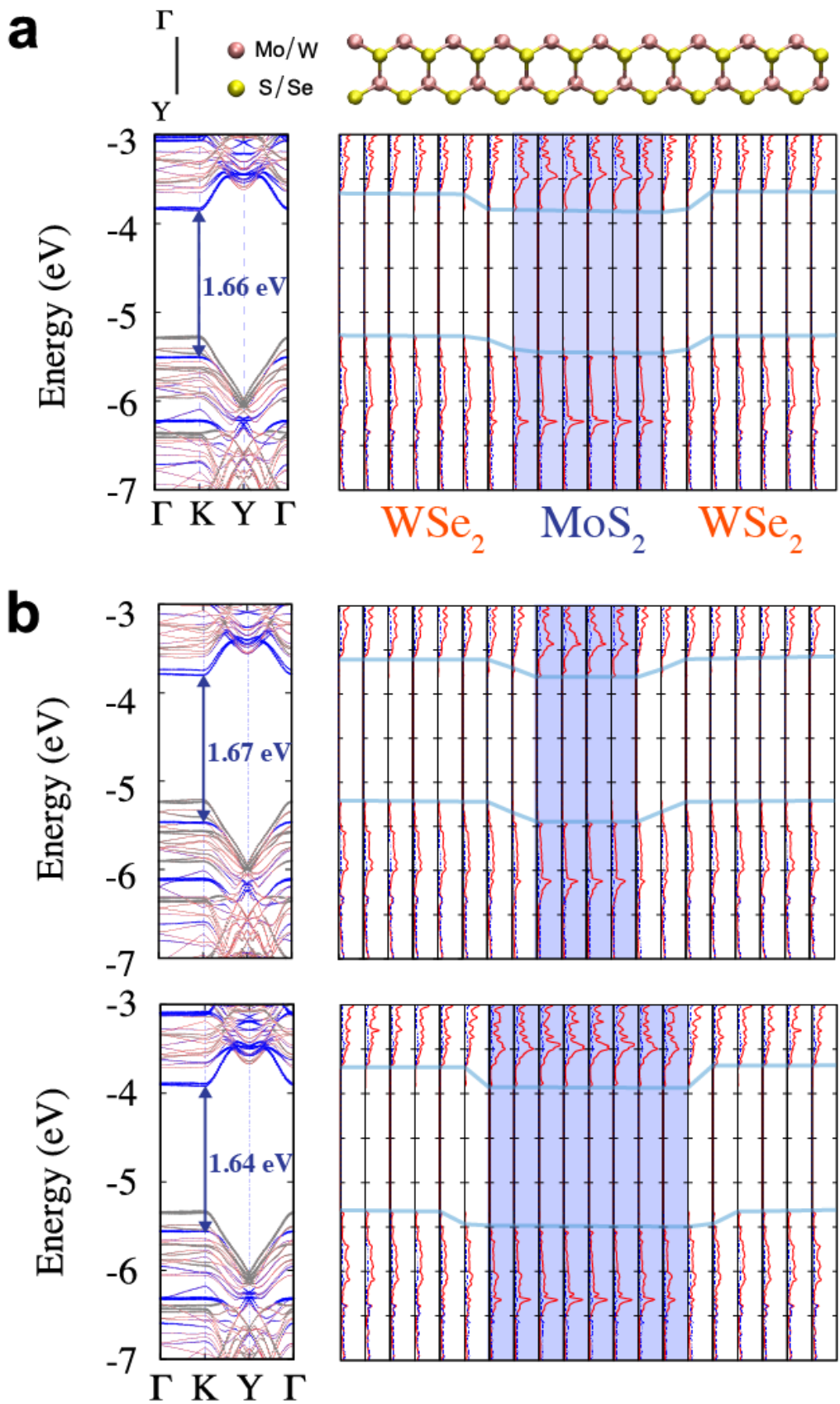

**Supplementary Figure 12 | Electronic properties of 1D MoS₂ channels embedded within WSe₂ monolayer. a,** DFT calculation of the orbital projected band structure

(left) and the projected density of state (PDOS) (right) of three-unit-cell wide $MoS_2$ 1D channel embedded within $WSe_2$. In the orbital projected band structures, the blue dot lines are corresponding to the weighted contribution from 1D $MoS_2$, where the dot size is proportional to the $MoS_2$ contribution. Our calculations show our 1D $MoS_2$ forms a direct band gap, which is different from the indirect band gap of uniaxially strained $MoS_2$ that has been reported before (ref. 22 in main manuscript). The gray dot lines are corresponding to the weighted contribution from the $WSe_2$, while the red lines show the total band structure. Moreover, the PDOS plots show clear type II band alignment (blue lines) that can potentially be used for charge separation. **b,** DFT calculations of 1D $MoS_2$ channels with two-unit-cell and four-unit-cell width. The band structure and PDOS show little difference from the three-unit-cell 1D channel, indicating robust 1D confinement even in the presence of small width variations. We chose the vacuum level as reference (0 eV). The spin-orbital coupling is not considered in this calculation.

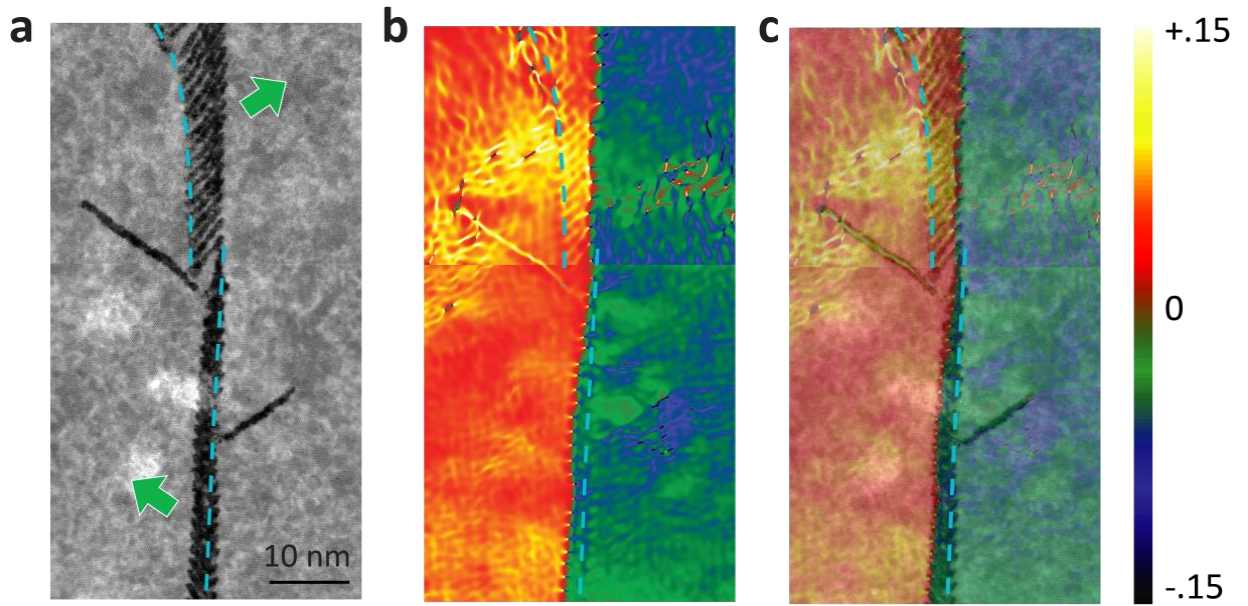

**Supplementary Figure 13 | Generation of superlattices. a,** Magnified ADF-STEM image from Fig. 4b with the original WSe$_2$ grain boundary marked with blue dashed lines. The green arrows indicate the collective migration direction of dislocations. Since the original grain boundary is not a straight line, the dislocations from the top and bottom parts migrate towards different directions to form a new straight grain boundary, where all dislocations tend to lie vertically above one another to achieve the lowest energy. **b,** The rotation map of **a**, showing the new straight grain boundary. **c,** The overlay of **a** and **b**.

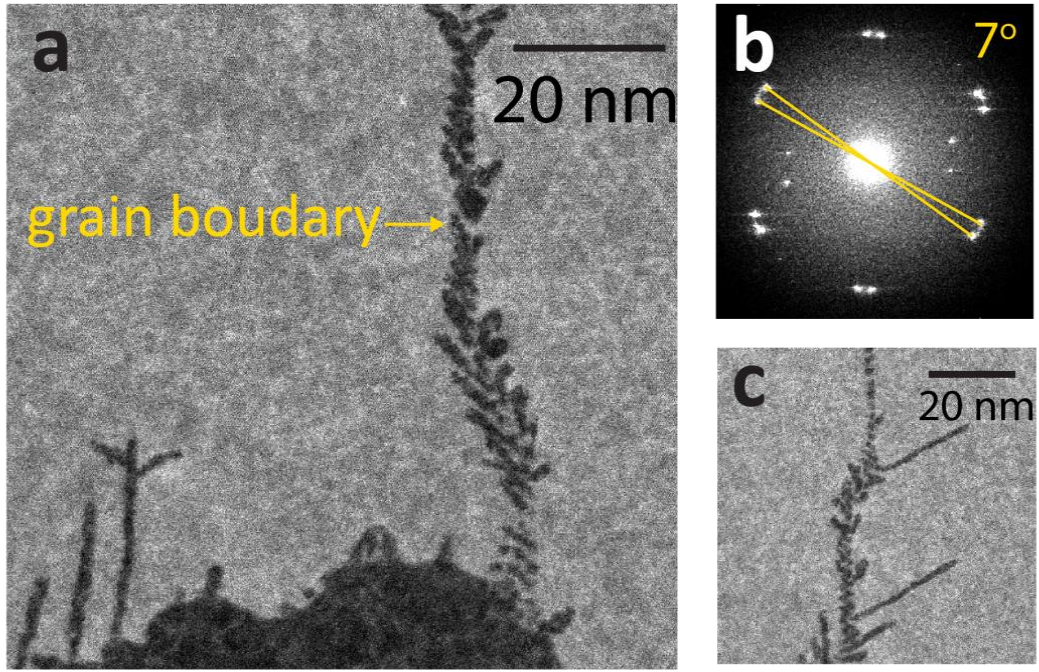

**Supplementary Figure 14 | MoS$_2$ superlattice forming at a 7° grain boundary in WSe$_2$. a,** ADF-STEM image of the MoS$_2$ superlattice forming at a 7° grain boundary. **b,** The diffractogram with the twisted angle measured. **c,** Another region of the grain boundary. The low-angle grain boundaries are lines of 5|7 dislocation arrays with the spacing following the classic dislocation theory $d = b/\theta$, where b is the Burger's vector and $\theta$ is the tilt angle. For example, at this 7° (0.12 in radians) grain boundary, the dislocation spacing should be 2.77 nm.

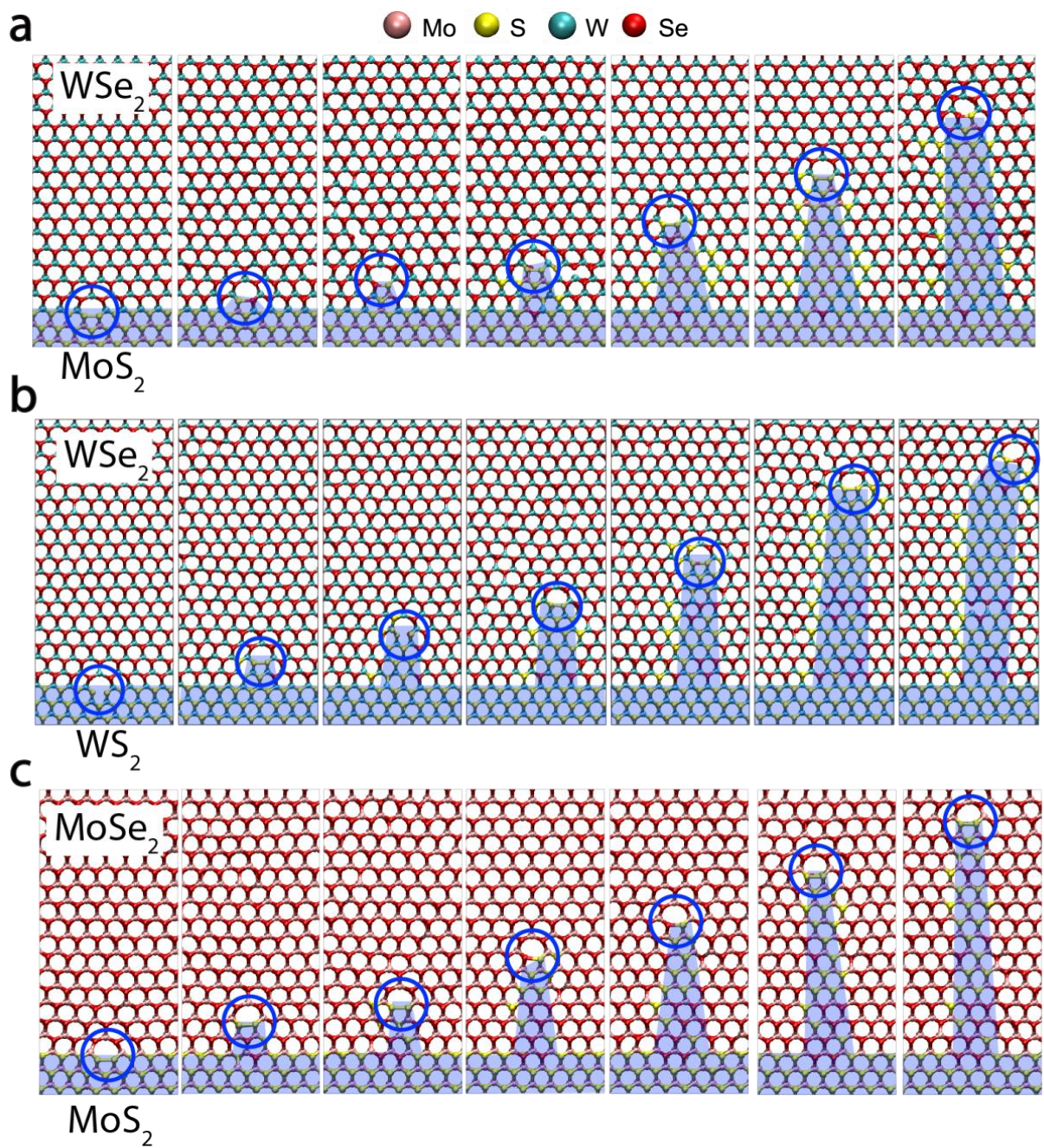

**Supplementary Figure 15 | MD simulations of different combinations of TMDs that grow 1D channels. a-c,** MD simulation of the formation of 1D $MoS_2$ embedded within 2D $WSe_2$ (**a**), 1D $WS_2$ embedded within 2D $WSe_2$ (**b**), and 1D $MoS_2$ embedded within 2D $MoSe_2$ (**c**). The results prove that this approach can extend to different combinations of TMDs other than $MoS_2$-$WSe_2$, although the widths of the 1D channels show small variations, which is due to the difference in optimized growth conditions for different cases.

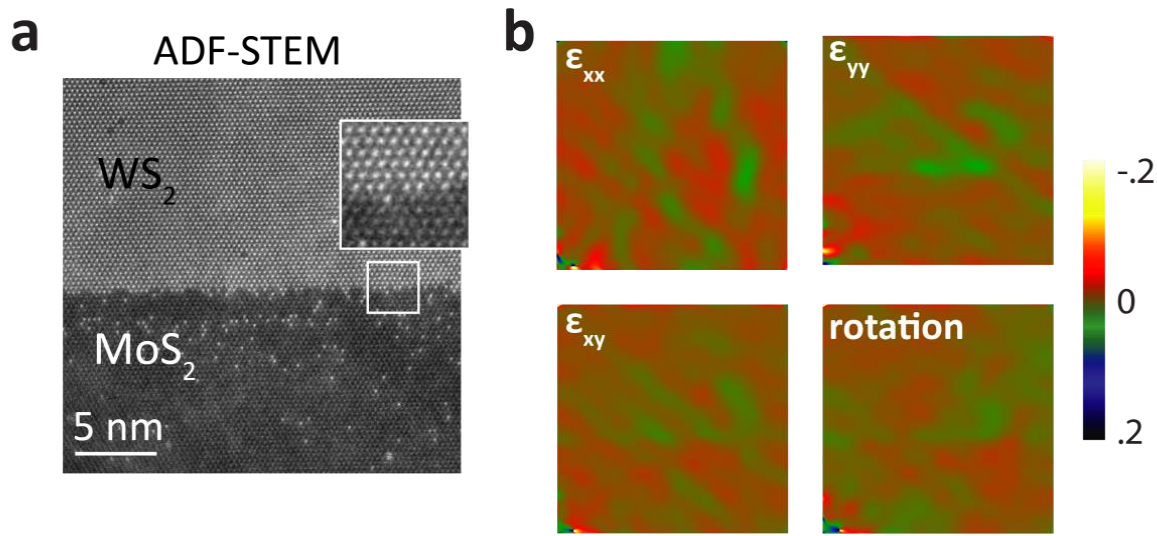

**Supplementary Figure 16 | Lack of 1D channels between MoS$_2$-WS$_2$. a,** Atomic-resolution ADF-STEM image of the MoS$_2$-WS$_2$ lateral heterojunction interface. **b,** The GPA maps of **a**, showing no strain, dislocations, or 1D channels forming at the MoS$_2$-WS$_2$ junction because the two materials are originally lattice matched.

**Supplementary Discussion 1: Development of reactive force field for molecular dynamics.**

Molecular Dynamics (MD) simulations in this study were performed via LAMMPS MD package[1] using the Reactive Empirical Bond Order (REBO) force field[2-4] to model the interactions among Mo-W-S-Se atoms in the atomically sharp $MoS_2$-$WSe_2$, $WS_2$-$WSe_2$, and $MoS_2$-$MoSe_2$ lateral heterojunctions. We utilized the optimized version of Mo-S REBO for $MoS_2$ monolayer by adjusting radius cutoffs and rescaling the attractive and repulsive terms of the original form[3,4], which have successfully described the experimental observation of $MoS_2$ monolayer crack propagation with vacancies of sulfides[5]. We developed W-Se, W-S, and Mo-Se REBO with the same bond-order parameters of Mo and S but different ionic radiuses on the basis of their chemical and structural similarities. The REBO force field has the form

$$E_b = \frac{1}{2} \sum_{i \neq j} f_{ij}^C(r_{ij}) \left[ \left(1 + \frac{Q}{r_{ij}}\right) A e^{-a r_{ij}} - b_{ij} B e^{-b r_{ij}} \right], \quad (1)$$

where $E_b$ is the binding energy, $f_{ij}^C(r_{ij})$ is a switching function, $b_{ij}$ is bond-order term, and $r_{ij}$ is the interatomic distance between atom $i$ and $j$. We note that $Q$, $a$ and $b$, are related to geometries such as equilibrium distances, A and B are related to the energy of attractive and repulsive terms. The radius cutoffs of the switching functions are critical to realistic bond breaking and forming because the functions can cause artificial forces near the failure points[5-7].

We obtained the geometric parameters for $MoS_2$, $MoSe_2$, $WS_2$, and $WSe_2$ from our Density Functional Theory (DFT) calculations by Quantum-Espresso package[8] using Perdew Burke Ernzerhof (PBE) functional[9] and norm-conserving type pseudopotential[10,11]. We prepare a rectangular shape cell containing 6 atoms with the periodic boundary condition in $x$ (along the zigzag edge) and $y$ (along the armchair edge) directions. To model each 2D material, the vacuum space of 15 Å in the $z$ direction is inserted to avoid unphysical interactions between periodic images. The energy cutoff for the wave functions is 60 Ry and $11 \times 11 \times 1$ grids are adopted for the K space sampling. Table S1 shows the results of the geometric parameters of four different monolayers.

The equilibrium distances between Mo-S and W-S are very similar (2.09 and 2.08 Å from our DFT calculations, respectively). In the REBO forms for $MoS_2$, the equilibrium distance between sulfides does not affect the lattice constants of monolayers. The difference of lattice parameters between $WS_2$ and $MoS_2$ mainly comes from the larger ionic radius of W than Mo. Thus, we parameterized $Q$, $a$ and $b$ in Eq. (5) of W-W to fit $WS_2$ lattice constants, while for the FF parameters of W-S we use the same $Q$, $a$ and $b$ of Mo-S.

Based on the obtained new parameters of W-W, $Q$, $a$ and $b$ of W-Se are parameterized to match the lattice parameters of $WSe_2$. In the same way, $Q$, $a$ and $b$ of Mo-Se are parameterized. We used $Q$, $a$ and $b$ of S-S for the parameters of Se-Se because Se-Se/S-S interaction are not important for the lattice constants of the monolayers. Table S2 shows the geometric parameters obtained from new REBO force field, which describes the difference between four different monolayers well.

We re-parameterized $A$ and $B$, which are related to repulsive and attractive terms. We followed the same strategy of the previous study[7], adjusting radius cut-offs, and rescaling A and B simultaneously to match stress-strain curves of monolayers, which were obtained from our DFT calculations (All conditions are the same as those for geometric parameters). We adjusted the radius cutoffs of the switching functions, $f_{ij}^C(r_{ij})$, to match the failure strains, and rescaled $A$ and $B$ in Eq. (5) to match the stresses at 0.1 strains in the y direction. The elastic constants from DFT and REBO are shown in Table S3 and S4. The new parameters well describe the relative differences of four different monolayers.

Finally, we extended the code from handling two atom types (Mo-S) to four different atom types (Mo-S-W-Se). For Mo-W and S-Se interactions, we used Tersoff potential mixing rules[12] for $a$, $b$, $A$, $B$ and radius cutoffs of the switching functions.

| DFT (Å) | $MoS_2$ | $WS_2$ | $MoSe_2$ | $WSe_2$ |
|---|---|---|---|---|
| Mo/W-Mo/W | 3.18 | 3.23 | 3.32 | 3.36 |
| Mo/W-S/Se | 2.45 | 2.45 | 2.55 | 2.58 |
| S/Se-S/Se | 3.18 | 3.23 | 3.32 | 3.36 |
| t(top-bottom) | 3.20 | 3.19 | 3.35 | 3.40 |

**Supplementary Table 1.** The geometric parameters obtained from DFT calculations.

| REBO (Å) | $MoS_2$ | $WS_2$ | $MoSe_2$ | $WSe_2$ |
|---|---|---|---|---|
| Mo/W-Mo/W | 3.17 | 3.24 | 3.28 | 3.33 |
| Mo/W-S/Se | 2.46 | 2.46 | 2.55 | 2.56 |
| S/Se-S/Se | 3.17 | 3.24 | 3.28 | 3.33 |
| t(top-bottom) | 3.23 | 3.21 | 3.41 | 3.38 |

**Supplementary Table 2.** The geometric parameters obtained from REBO potential

| DFT (N/m) | $MoS_2$ | $WS_2$ | $MoSe_2$ | $WSe_2$ |
|---|---|---|---|---|
| C11 | 129.9 | 129.3 | 104.7 | 113.1 |
| C22 | 130.2 | 132.6 | 105.6 | 112.2 |
| C12 | 29.25 | 30.15 | 24.9 | 25.8 |
| E | 123.5 | 124 | 99.2 | 106.7 |

**Supplementary Table 3.** Elastic constants and Young's modulus (averaged in the *x* and *y* directions) from DFT calculations

| REBO (N/m) | $MoS_2$ | $WS_2$ | $MoSe_2$ | $WSe_2$ |
|---|---|---|---|---|
| C11 | 115.1 | 118.3 | 90.1 | 96.3 |
| C22 | 115.1 | 118.3 | 90.1 | 96.3 |
| C12 | 34.1 | 34.8 | 26.7 | 28.6 |
| E | 105.0 | 108.1 | 82.2 | 87.8 |

**Supplementary Table 4**. Elastic constants and Young's modulus (averaged in the *x* and *y* directions) from the current REBO potential

**Supplementary Discussion 3: Molecular dynamics model for 1D channels.**

We utilized our new REBO force fields for 1D MoS$_2$ channel evolution study. The computational model is a heterojunction composed of MoS$_2$ (4 nm x 7 nm) and WSe$_2$ (6 nm x 7 nm) with an interface along the zigzag edge of the 2D lattice. The 7 nm as the length of the material interface of this heterojunction is naturally given by the ~5% lattice mismatch between MoS$_2$ and WSe$_2$ (Supplementary Table 1). Periodic boundary conditions were applied to the zigzag direction along the interface of the heterojunction. Perpendicular to the interface, the model has 2 nm spacing between the simulation box boundary and the MoS$_2$ edge, and 2 nm spacing for the WSe$_2$ edge from the simulation box boundary. In addition, we set a 10 nm void region in the out-of-plane direction of the 2D material. These margins are large enough to guarantee that MoS$_2$ and WSe$_2$ only interact at the interface of the heterojunction.

We applied an interaction between the bottom layer of S/Se atoms and the substrate by using a Lennard-Jones (LJ) intermolecular potential with the 9-3 form[13] of

$$E_{\text{Sub}} = \epsilon \left[ \frac{2}{15} \left( \frac{\sigma}{r} \right)^9 - \left( \frac{\sigma}{r} \right)^3 \right], \tag{2}$$

where $r$ is the distance from an atom to the surface of substrate; $\sigma$ and $\epsilon$ are parameters that relates to the equilibrium distance $r = 0.858\sigma$ and adhesion energy $1.054\epsilon$ per atom. The form is related to the integration over a half-lattice of particles with LJ 12-6 intermolecular potential. In the simulation, we used $\sigma$=2.3 Å and $\epsilon =$ 0.1 eV through our simulations. We chose the adhesion energy as the system is stabilized during the annealing process (See more details in Supplementary discussion 4). We note that the substrate model is simplified as an infinite wall to prevent the out-of-plane deformation and the penetration of atoms, which is similar to the sapphire substrate we used in experiments.

We applied a number of cyclic annealing processes to study the behaviors of the dislocations. As a result, the pentagon-heptagon dislocations clearly climb towards the heptagon direction after hundreds of iterations as shown in Supplementary Fig. 15 and Supplementary movie #9, which matches our experimental results and represents the dynamic process for 1D MoS$_2$ formation. The supplementary movie #10 and #11 show the growth for 1D WS$_2$ in WSe$_2$ and 1D MoS$_2$ in MoSe$_2$, indicating that this approach can be extended to other combinations of transition metal dichalcogenides (TMDs).

**Supplementary Discussion 4: Annealing process for structural evolution.**

A number of cyclic annealing processes with adding or deleting atoms near the dislocation equilibrate the MoS$_2$-WSe$_2$ model, which is an algorithm combining Monte Carlo and Molecular Dynamics to accelerate the evolutions of the structures. Each of the cycles is composed of 4 stages by using NPT or NVT ensemble: heating, relaxing at high temperature, cooling, and relaxing at low temperature. The periodic condition and NPT ensemble are applied during the 2$^{nd}$ stage (relaxing at high temperature) to allow the structural relaxation along the junction direction. We have $T_{low}$=600 K for all atoms, and $T_{high}$=1100 K for W, $T_{high}$=900 K for Mo and Se, and $T_{high}$=600 K for S with 25 ps for the 2$^{nd}$ stage and 5 ps for the others. Before the 1$^{st}$ stage (heating) and after the 4$^{th}$ stage (relaxing at low temperature), conjugate gradient minimizations are applied for 5,000 steps. To simulate the experimental process of depositing Mo and S precursors at high temperature, we add MoS$_2$ nanoparticles 3 Å above the Mo/W plane over the dislocation with random variations less than 1.0 Å in both lattice directions, which allows both simulation efficiency and natural reaction between the nanoparticles and the MoS$_2$-WSe$_2$ heterojunction to be possible.

After every annealing process, we estimate a position of the next Mo for the most spacious region. When an estimated Mo position is located 2.4 Å away from the Mo/W atoms, the precursor nanoparticle (MoS$_2$) is added on the top of the system for the next cycle. The S$_2$ or S$_4$ is added based on the total number of sulfur and selenium atoms. After the 2$^{nd}$ stage, we check the out-of-plane displacements of atoms, and delete S/Se above 3.0 Å and Mo/W above 1.5 Å away from the Mo/W plane. We note that all these processes are setup to accelerate the reactions without forcibly forming bonds or other structures. The result confirms the generation of a straight 1D channel from the catalyst 5|7 dislocation with atomic thickness and reveals the physics behind the growing process in atomic scale. To predict whether the approach can be applied to other TMDs, similar annealing process was conducted during the simulation in WS$_2$-WSe$_2$ and MoS$_2$-MoSe$_2$ as well.

**Supplementary Discussion 5: 1D MoS$_2$ channels step-by-step formation**

Supplementary Movie #1 shows how the precursor nanoparticle (MoS$_2$) around the 5|7 dislocation inserts into the lattice and forms an intermediate state (Fig. 3a in the main manuscript). Although the metal precursor comes with sulfur atoms around, in some cases the sulfurs are released back to the environment due to the lack of the preferable condition. The reaction can occur with different number of sulfur atoms (movie #2: MoS, movie #3: MoS$_3$, movie#4: MoS$_4$), but it does not occur without sulfur due to the low coordination number of molybdenum atom (the Mo bonded to four local S or Se). In this case, the injection is forbidden, ending with the Mo taking local S or Se out of the system. This is a very localized reduction around the dislocation region, indicating that the catalyst dislocation cannot grow 1D channel with little sulfur precursor in the system (movie #5). Moreover, the precursors away from the catalyst misfit dislocation can't substitute or insert into the original lattice, indicating the highly confinement of this dislocation catalyzed approach (movie #6).

The injection of the metal atom still leaves space for the next coming sulfur atoms to enter and bond to metal atoms, which may involve breaking and reforming bonds. As a result, the misfit dislocation climbs. Supplementary movie #7&8 show the top and side view of how the dislocation climbs up, trying to form the next 5|7 dislocation. In these movies (#7&8), due to insufficient S in the system, the atoms near the dislocation move actively to find energetically favorable configuration. During this process, some atoms near the catalyst dislocation were pushed out of plane and left the system, and the newly introduced molybdenum or sulfur atoms sometimes substitute the original tungsten or selenium. We note only atoms near dislocation can be substituted.

We also found that the ratio between the precursors (S:Mo) is critical to stabilize the growth in MD simulation and form perfect 5|7 dislocation, which is consistent with the growth conditions in experiments. Sufficient sulfur should be provided to form perfect 5|7 dislocation for the stable and long growth. However, too much sulfur would take too much pre-existing metal atoms out before forming the next dislocation. We also found the same mechanisms with other combination of TMDs, such as 1D WS$_2$ channel embedded in WSe$_2$ and 1D MoS$_2$ embedded in MoSe$_2$ (Movie #10&11).

We summarized and simplified the entire process into four main steps:

1) Insertion of Mo precursor atoms at the 5|7 catalyst dislocation.

2) Reconstruction and relaxation of the structure with more S precursor atoms to form the next 5|7 dislocation.

3) Repeat the process to form 1D channels.

**Supplementary discussion 6: DFT calculation of 1D MoS$_2$ channel embedded in WSe$_2$**

To demonstrate possible applications of 1D MoS$_2$ channel embedded in WSe$_2$, we obtained orbital projected band structures and projected density of state (PDOS) of six different models by QE package using the same functional and pseudopotential used for geometric parameters in Supplementary Discussion 2. The rectangular shape unit cells in replicated in the x direction containing 60 atoms with the periodic boundary condition in *x* (along the zigzag edge) and *y* (along the armchair edge) directions. The energy cutoff for the wave functions is set to 60 Ry and 2×8×1 and 4x32x1 Monkhost-Pack grids are adopted for the K space sampling for structure relaxation and PDOS, respectively. We prepare 6 models with different ratios of $MoS_2$ and $WSe_2$ to represent 1D MoS2 channels. The cells and atomistic structures are fully relaxed with convergence thresholds of 0.5 kbar and $10^{-3}$ (a.u.) for the pressure and atomic forces, respectively. Due to the rectangular unit shape of the system, we used Γ-K-Y-Γ for the bands structure to see the difference between the gap at the Γ the K points in hexagonal Brillouin zone (BZ) as suggested in the previous study[14]. Absolute conduction band minimum (CBM) and valence band maximum (VBM) relative to the vacuum level are calculated for all models. The obtained energy levels of pristine $MoS_2$ and $WSe_2$ show good agreement with those from the previous study[15], confirming the reliability of our calculation.

From the orbital projected band structures and PDOS shown in Supplementary Fig 16, the 1D channel of $MoS_2$ mainly contributes to the CBM and the major contribution for the VBM comes from $WSe_2$. To estimate the local band gaps of $MoS_2$, we evaluated the contributions of $MoS_2$ for each eigenvalue in the band structures from the orbital projected band structure. As shown in Supplementary Fig 16, the blue dot lines are corresponding to the weighted contribution from 1D $MoS_2$, where the dot size is proportional to the $MoS_2$ contribution. The DFT calculations show that the gap of 1D $MoS_2$ is direct band gap even if we consider both K and Γ points while the uniaxial strain applied to $MoS_2$ results in the transition from the direct to indirect band gap. Due to the 1D confinement, the uniqueness of the band structure of 1D $MoS_2$ becomes more distinct, considering only K-Y or Γ-Y path in the BZ, where K-Γ path is ignored due to the translational symmetry broken in that path. The detailed values are listed in Supplementary Table 6.

|  | $(MoS_2)_{10}$ | $(WSe_2)_{10}$ | $(MoS_2)_1$ $(WSe_2)_9$ | $(MoS_2)_2$ $(WSe_2)_8$ | $(MoS_2)_3$ $(WSe_2)_7$ | $(MoS_2)_4$ $(WSe_2)_6$ |
|---|---|---|---|---|---|---|
| A in x (Å) | 31.86 | 33.60 | 33.43 | 33.24 | 33.07 | 32.90 |
| B in y (Å) | 5.52 | 5.82 | 5.79 | 5.76 | 5.73 | 5.70 |
| $\varepsilon_y$ in MoS$_2$ | 0 | - | 4.9% | 4.3% | 3.8% | 3.2% |
| $\varepsilon_y$ in WSe$_2$ | - | 0 | -0.5% | -1.0% | -1.5% | -2.0% |
| CBM (eV) | -4.21 (-4.29)[17] | -3.64 (-3.69)[17] | -3.72 | -3.8 | -3.85 | -3.92 |
| VBM (eV) | -5.92 (-5.98)[17] | -5.09 (-5.20)[17] | -5.15 | -5.21 | -5.27 | -5.33 |
| Gap (eV) | 1.71 (1.69)[17] | 1.45 (1.51)[17] | 1.43 | 1.41 | 1.42 | 1.41 |

**Supplementary Table 5**. The relaxed lattice parameters (A in the x direction and B in the y direction), CBM, VBM and band gap of each model. The reference values of pristine MoS$_2$ and WSe$_2$ are obtained from the previous study[15]. Here we use the foot notation to indicate the width of the materials in rectangular unit cells. For example, (MoS$_2$)$_3$(WSe$_2$)$_7$ indicate a three-unit-cell wide MoS$_2$ 1D channel embedded within a seven-unit-cell WSe$_2$ matrix, which is schematically described in Supplementary Fig. 16a.

| | (MoS$_2$)$_1$ (WSe$_2$)$_9$ | (MoS$_2$)$_2$ (WSe$_2$)$_8$ | (MoS$_2$)$_3$ (WSe$_2$)$_7$ | (MoS$_2$)$_4$ (WSe$_2$)$_6$ |
|---|---|---|---|---|
| CBM@MoS$_2$ (eV) | -3.72 | -3.80 | -3.85 | -3.92 |
| VBM@MoS$_2$ (eV) | -5.24 | -5.47 | -5.51 | -5.56 |
| Gap@MoS$_2$ (eV) | 1.52 | 1.67 | 1.66 | 1.64 |

**Supplementary Table 6**. The estimation of localized band-edge of MoS$_2$ from the orbital projected band structures. With widths from one rectangular unit cell to four unit cells, the MoS$_2$ channels all present direct band gaps even if we consider the $\Gamma$ point in the calculation.